\documentclass[twocolumn]{article}
\usepackage{graphicx,comment,amsmath} 

\begin{document}

\title{Principles and design of a Zeeman-Sisyphus decelerator for molecular beams}

\date{\today}

\author{N. J. Fitch \thanks{Centre for Cold Matter, Blackett Laboratory, Imperial College London, Prince Consort Road, London SW7 2AZ, UK} \and M. R. Tarbutt\footnotemark[1]}

\maketitle

\section{Abstract}
We explore a technique for decelerating molecules using a static magnetic field and optical pumping.  Molecules travel through a spatially varying magnetic field and are repeatedly pumped into a weak-field seeking state as they move towards each strong field region, and into a strong-field seeking state as they move towards weak field. The method is time-independent and so is suitable for decelerating both pulsed and continuous molecular beams. By using guiding magnets at each weak field region, the beam can be simultaneously guided and decelerated. By tapering the magnetic field strength in the strong field regions, and exploiting the Doppler shift, the velocity distribution can be compressed during deceleration. We develop the principles of this deceleration technique, provide a realistic design, use numerical simulations to evaluate its performance for a beam of CaF, and compare this performance to other deceleration methods.

\section{Introduction}
\label{Introduction}

There has been much recent progress in the formation and control of cold molecules, motivated by numerous potential applications~\cite{Krems,Carr2009}, including quantum information processing~\cite{DeMille2002}, tests of fundamental physics~\cite{Hudson2006,DeMille2008,Isaev2010,Hudson2011,Truppe2013,Eckel2013,Baron2014}, and understanding chemistry and intermolecular collisions at the quantum level~\cite{Lee1987,Casavecchia2000,Krems2008,Chandler2010}.  With many experiments beginning with a relatively fast molecular beam, deceleration techniques such as Stark deceleration~\cite{Bethlem1999}, Zeeman deceleration~\cite{Narevicius2008}, optical Stark deceleration~\cite{Fulton2004} and centrifuge deceleration~\cite{Chervenkov2014}, have been at the forefront of cold molecule research, being used to provide velocity-controlled beams and to load traps that can store cold molecules for many seconds~\cite{Bethlem2000}.  Once trapped, a common goal is to cool the molecules to lower temperatures by sympathetic~\cite{Tokunaga2011, Lim2015}, Sisyphus~\cite{Zeppenfeld2009, Zeppenfeld2012, Prehn2016, Comparat2014}, adiabatic~\cite{Perez2013} or evaporative~\cite{Stuhl2012} cooling.  Direct laser slowing and cooling is another viable option for certain species~\cite{Shuman2010,Hummon2013,Zhelyazkova2014,Truppe2016}, being capable of both decelerating and subsequently trapping and cooling the molecules under study.  Indeed, such an approach has recently led to the demonstration~\cite{Barry2014} and optimization~\cite{McCarron2015,Norrgard2016} of the first molecular magneto-optical trap (MOT).
In this experiment, a cryogenic buffer-gas source produces intense molecular pulses, typically 1--10~ms in duration with speeds in the range 50--200~m/s, depending on the source geometry and gas flow rate~\cite{Maxwell2005,Hutzler2012}. Then, by scattering about $10^{4}$ photons from a counter-propagating laser beam~\cite{Barry2012b}, the molecules are decelerated to the capture velocity of the MOT, which is about 10~m/s~\cite{Tarbutt2015}. So far, only a few thousand molecules have been captured, mainly because the slowing method is inefficient. There are several reasons for this: (i) the stopping distance is large compared to the capture area of the trap, and so the solid-angle that can be captured is small; (ii) the molecular beam is slowed longitudinally, but is not cooled transversely, and so the beam divergence grows as the molecules are slowed; (iii) the photon scattering that slows down the beam also causes transverse heating, which increases the divergence even further; (iv) molecules are lost if they decay out of the cooling cycle, and addressing those decays increases the experimental complexity.

A current focus of research in this area is to increase the number of molecules loaded into MOTs by improving the efficiency of the deceleration process.  Such progress is important both for current experiments and to extend laser slowing and cooling to diatomic and polyatomic~\cite{Kozyryev2016} species with less favorable vibrational branching ratios.  Decelerators that use time-dependent fields, such as Stark decelerators, are not well suited to this application because they slow a few slices of the molecular beam that are only a few mm in length, a hundred times shorter than the beams emitted by a typical buffer-gas source. A traveling-wave decelerator~\cite{Fabrikant2014} or centrifuge decelerator~\cite{Chervenkov2014} can handle long pulses, but these methods have not yet been widely adopted. DeMille {\it et al.}~\cite{DeMille2013} explore methods to confine a molecular beam transversely as it is slowed by radiation pressure, and conclude that guiding using microwave fields is a good option.

Here, we explore a technique which we call Zeeman-Sisyphus deceleration. Molecules in a beam travel through an array of permanent magnets that produces a spatially varying magnetic field, and are optically pumped into a weak-field seeking state as they move towards regions of strong field, and into a strong-field seeking state as they move towards regions of weak field. In this way, there is always a force opposing their forward motion. This general idea has a long history. In 1981, Breeden and Metcalf suggested a similar method for decelerating atoms in Rydberg states~\cite{Breeden1981}.  More recently, the use of Sisyphus-type forces due to the Stark effect has been proposed as a deceleration technique~\cite{Hudson2009} and demonstrated with great success as a cooling method for electrostatically trapped molecules~\cite{Zeppenfeld2012,Prehn2016}.  The magnetic-field analogs, relying on the Zeeman effect, have also been discussed~\cite{Comparat2014} and already form the basis of an established trap-loading technique~\cite{Riedel2011,Lu2014}.  Here, we analyze in detail the prospects for extending these techniques to molecular beam deceleration, finding a design that provides both longitudinal slowing and net transverse guiding, as required of a viable deceleration method.  The approach is capable of bringing a typical buffer-gas-cooled molecular beam to rest by scattering only a few hundred photons, far less than the typical $\sim 10^{4}$ scattered photons required for direct laser slowing, and so could be applied to molecular species with only quasi-closed optical cycling transitions without needing numerous repump lasers.

\section{Principles}
\label{Zeeman_Sisyphus_Deceleration}

\begin{figure}[!htb]
\centering
\includegraphics[width=\linewidth]{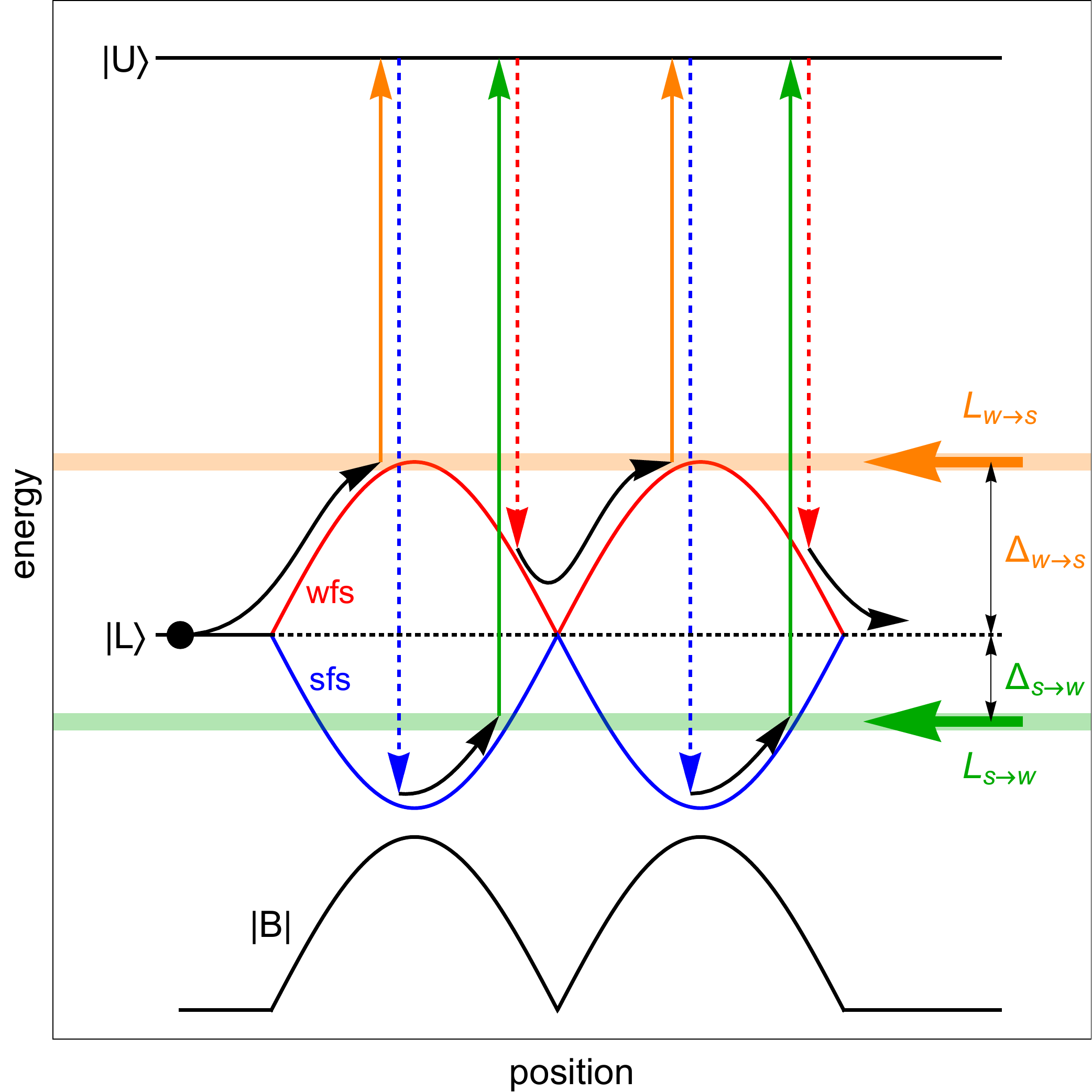}
\caption{Illustration of the Zeeman-Sisyphus deceleration technique.  A ground state molecule (black dot) propagates through two regions of large magnetic field (two hills/valleys in energy) and is periodically optically pumped between weak-field and strong-field seeking states such that it is perpetually decelerated.}
\label{basicIdeaFig}
\end{figure}

The general idea of Zeeman-Sisyphus deceleration is illustrated in Figure~\ref{basicIdeaFig}.  Here, a molecule (black dot) with a lower state $L$ and upper state $U$ propagates through two regions of strong magnetic field separated by a region of weak field.  $L$ is degenerate in zero field with two substates which shift oppositely in the applied field: a weak-field seeking (wfs) state whose energy increases with field strength, and a strong-field seeking (sfs) state whose energy decreases with field strength. The upper state has no Zeeman shift, and it can decay to either of the two lower states, but not to any other state of the molecule. Molecules that are amenable to laser cooling approach this ideal, while for others there may be transitions to other rotational or vibrational states which would need to be addressed. The two lower states are coupled to the upper state by two pump lasers, $L_{w \rightarrow s}$ and $L_{s \rightarrow w}$. $L_{w \rightarrow s}$ has a negative detuning of $-\Delta_{w \rightarrow s}$ relative to the zero-field resonance frequency, while $L_{s \rightarrow w}$ has a positive detuning of $\Delta_{s \rightarrow w}$. The magnetic field values at which the molecules come into resonance with one of the lasers are called the {\it resonance fields}, and the locations in space where this occurs are called the {\it resonance points}. Both $\Delta_{w \rightarrow s}$ and $\Delta_{s \rightarrow w}$ are positive quantities, and they are arranged with $\Delta_{w \rightarrow s} > \Delta_{s \rightarrow w}$. With this configuration, wfs (sfs) molecules come into resonance with $L_{w \rightarrow s}$ ($L_{s \rightarrow w}$) in regions of strong (weak) magnetic field and are then optically pumped to the other state by absorption and subsequent spontaneous decay, as indicated by the solid and dashed vertical arrows, respectively.  The distance moved by the molecules during the optical pumping process is negligible. The molecules decelerate because they move into each strong field region in a weak-field seeking state, and out of those regions in a strong-field seeking state. This process is repeated until the molecules reach the desired final velocity.

For the arrangement shown in Figure~\ref{basicIdeaFig}, the average deceleration force is
\begin{equation}
\label{deceleration}
\bar{F}_{z} = \frac{h \left(\Delta_{w \rightarrow s} - \Delta_{s \rightarrow w} \right)}{L},
\end{equation}
where $2L$ is the spatial periodicity and the laser detunings are given in Hz.  For the largest force we should set $h \Delta_{w \rightarrow s} = U_{\rm{max}}$, where $U_{\rm{max}}$ is the maximum Zeeman shift, and $\Delta_{s \rightarrow w}$=0.  However, to ensure that $L_{s \rightarrow w}$ only pumps molecules out of strong-field seeking states, $\Delta_{s \rightarrow w}$ should not be too close to zero. For a fixed decelerator length, the change in speed due to the average constant force of equation~(\ref{deceleration}) is inversely proportional to the mean speed, and so deceleration increases the spread of velocities in the beam. In Sec.~\ref{Advanced_Methods} we show how the Doppler shift can be used to counter this effect under certain conditions.

Because the deceleration method is time-independent, it is applicable to long-pulse or even continuous molecular beams. It works for molecules of all longitudinal positions and speeds, and so its longitudinal phase-space acceptance is unbounded. We would also like to arrange a large transverse phase-space acceptance, meaning that molecules should be guided as they are decelerated.  Because the strongest fields are at the magnet surfaces, molecules will tend to be anti-guided while in the strong-field seeking state and guided while in the weak-field seeking state.  With the arrangement of detunings illustrated in Figure~\ref{basicIdeaFig}, the molecules spend more of their time in the weak-field seeking state, and so net guiding seems possible. Moreover, we note that the molecules are in the weak-field seeking state as they pass through the field minimum, and that guiding magnets naturally have zero magnetic field on the axis. This presents an opportunity to interleave decelerating magnets, where the field is strong and uniform, with guiding magnets, where the field is weak and increases with transverse displacement from the axis. A suitable arrangement of permanent magnets that achieves this is presented in section \ref{Decelerator_Design}.

The deceleration method relies on efficient optical pumping of molecules between the weak- and strong-field seeking ground-states.  Molecules that are not optically pumped will not be decelerated as efficiently, and if they remain in a strong-field seeking state for too long their trajectories are likely to become transversely unstable.  To understand how the optical pumping efficiency depends on various experimental parameters, we introduce a simple analytical model of the optical pumping process.  Let $p$ be the probability that the molecule switches from one ground state to the other after scattering a single photon, and let $R(t)$ be the scattering rate at time $t$. The mean number of photons scattered by a molecule that is {\it not} optically pumped as it passes through the resonance point is $\bar{n}=\int R(t) \, dt$, where the integral is taken over the period of time where $R(t)$ is appreciable. In terms of $p$ and $\bar{n}$, the optical pumping probability $\chi$ is
\begin{equation}
\label{chi}
\chi = 1 - (1-p)^{\bar{n}}.
\end{equation}
For a two-level system, which is a reasonable approximation for our optical pumping arrangement, the steady-state scattering rate is~\cite{Metcalf}
\begin{equation}
\label{scattering_rate}
R = \frac{\Gamma}{2} \frac{s}{\left( 1 + s + 4 \, \delta^{2}/\Gamma^{2} \right)},
\end{equation}
where $\Gamma$ is the excited state decay rate, $s=I/I_{\rm{sat}}$ is the saturation parameter of the pump laser, and $\delta$ is the laser detuning,
\begin{equation}
\label{detuning}
\delta = 2 \pi \left(\Delta_{0} + \frac{U(|\vec{B}|)}{h} + \frac{v_{z}}{\lambda} \right).
\end{equation}
Here, $\lambda$ is the transition wavelength, $v_{z}$ is the forward velocity of the molecule, $-U$ is the Zeeman shift of the transition energy in a magnetic field $\vec{B}$, and $\Delta_{0}=f_{\textrm{laser}}-f_{0}$ is the detuning of the laser from the transition frequency for a stationary molecule in zero field. Over the small region of space around the resonance point where $R$ is large, $\delta$ changes approximately linearly with $z$ and hence with $t$, so we take $\delta =\beta t$. This gives
\begin{equation}
\bar{n} = \frac{s \Gamma}{2} \int_{-\infty}^{\infty} \frac{1}{1+s+4 \, (\beta t/\Gamma)^{2}} \, dt = \frac{\pi \, s \, \Gamma^{2}}{4 \, \sqrt{1+s} \, | \beta |}.
\end{equation}
Assuming a magnetic moment of $\mu_{B}$, and neglecting the very small change in speed as the molecule passes through the resonance point, we have
\begin{equation}
\label{beta}
\beta = \partial_{t}\delta \approx \frac{\mu_{B}}{\hbar} \partial_{t} B = \frac{\mu_{B} \, v_{z}}{\hbar} \partial_{z} B,
\end{equation}
where $\partial_{z} B$ is the longitudinal component of the gradient of the magnetic field magnitude at the resonance point.  Thus,
\begin{equation}
\label{PhotonsScattered}
\bar{n} = \frac{\pi \, \hbar \, s \, \Gamma^{2}}{4 \, \mu_{B} \, v_{z} \,\sqrt{1+s} \, | \partial_{z}B |}.
\end{equation}

Together, equations (\ref{chi}) and (\ref{PhotonsScattered}) determine the average optical pumping probability as a function of the relevant experimental parameters. This probability needs to be high enough to ensure that molecules pass through most of the guiding magnets in the weak-field seeking state, setting a requirement on $\bar{n}$. Re-arranging equation~(\ref{PhotonsScattered}) then gives a maximum allowable value for the field gradient at the resonance points. This maximum scales inversely with $v_z$, and scales linearly with $s$ when $s\ll$1 but only as $\sqrt{s}$ when $s\gg$1. The value of $p$ depends on the molecular transition and particular Zeeman sublevel, the magnitude of $\vec{B}$ at the resonance point, and the polarization of the optical pumping light relative to $\vec{B}$.  We investigate these details in Sections~\ref{CaF} and \ref{Trajectory_Simulations}, and find that a constant $p\approx$1/2 is a good approximation.

In this paper, we consider the prototypical case of decelerating CaF molecules emitted from a cryogenic buffer-gas source. The molecules are optically pumped on the A$^{2}\Pi_{1/2}$--X$^{2}\Sigma^{+}$ transition which has $\lambda$=606~nm, $\Gamma$=2$\pi\times$8.3~MHz, and $I_{\rm sat}$=5~mW/cm$^{2}$. A typical initial speed is $v_z$=150~m/s, and the corresponding kinetic energy is $h\times$1660~GHz. The molecules move through an array of permanent magnets that produce a peak field of $\simeq$1~T. The magnetic dipole moments of the ground states are $\pm \mu_{B}$, and so the maximum energy that can be removed per strong-field region, referred to as a deceleration {\it stage}, is $h\times$28~GHz. The minimum number of stages needed to bring the molecule to rest is 60.  Using $p$=1/2, the average number of photons scattered by decelerated molecules is 240, about 40 times smaller than using radiation pressure alone. Choosing $\bar{n}$=5 gives an optical pumping probability of approximately 97\%. A reasonable laser intensity is 250~mW/cm$^{2}$, corresponding to $s$=50, which gives a maximum allowable field gradient at the resonance points of about 2~T/cm. This sets an approximate scale of about 2~cm for the periodicity of the decelerator, giving an overall decelerator length of 1.2~m.

\section{Decelerator Design}
\label{Decelerator_Design}

\begin{figure}[!htb]
\centering
\includegraphics[width=\linewidth]{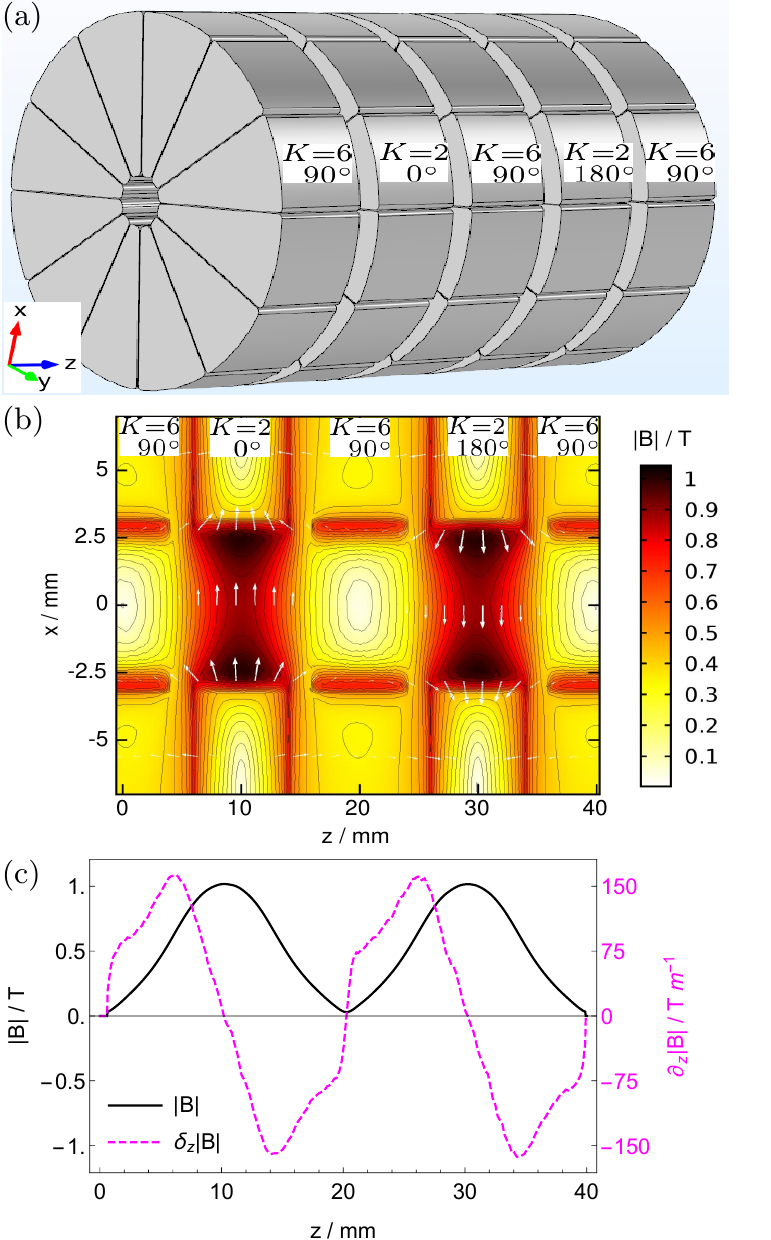}
\caption{Zeeman-Sisyphus decelerator design. (a) The magnet geometry consists of a stack of two types of approximate Halbach cylinders, denoted $K$=2 and $K$=6, which produce regions of strong and weak magnetic field, respectively. (b) A slice of the magnetic field magnitude as calculated by finite-element analysis methods. (c) The on-axis field magnitude (solid line) and its gradient (dashed line).}
\label{Fields}
\end{figure}

Figure~\ref{Fields}(a) illustrates our decelerator design, which follows the design principles outlined above.  It consists of an array of cylindrical permanent magnets whose axes are concentric with the molecular beam axis (z). The magnets alternate between two types of approximate Halbach cylinders~\cite{Halbach1980}, which are discussed in detail below.  The two types are denoted $K$=2 and $K$=6 in the figure with angle labels representing an absolute rotation relative to the global coordinate axes.  Each cylinder is 8~mm thick longitudinally with an outer diameter of 40~mm and an inner diameter of 5~mm through which the molecules propagate.  A 2~mm gap between the cylinders allows background gas to escape from the inner bore, and only slightly weakens the longitudinal magnetic-field gradient. Constructing this geometry out of N52 NdFeB wedge magnets with a remanent magnetization of 1.44~T results in the magnetic field shown in Figure~\ref{Fields}(b)-(c), calculated using finite-element methods.  Figure~\ref{Fields}(b) shows the magnetic field over a slice through the $xz$-plane.  The contours are lines of equal magnetic field magnitude and the white arrows show the field direction.  We see that the $K$=2 cylinders produce strong and fairly uniform magnetic fields while the $K$=6 cylinders provide regions of low magnetic field and transverse guiding of wfs molecules.  Note that every other $K$=2 cylinder is rotated 180$^{\circ}$ to produce strong-field directions that alternate between $\pm\hat{x}$ and fringe fields that cancel at the longitudinal centers of the $K$=6 stages.  Without this rotation the fringe fields from the strong-field stages produce an undesirable non-zero field offset in the guiding regions.  The orientation of the $K$=6 stages is chosen to give approximately equal field gradients in the two transverse directions.  Figure~\ref{Fields}(c) shows the magnetic field magnitude (solid line), and its gradient (dashed line), along the z-axis.  The on-axis field magnitude spatially oscillates between 0 and 1~T. The peak gradient is about 150~T/m, and so the condition on the maximum field gradient discussed in Sec.~\ref{Zeeman_Sisyphus_Deceleration} is satisfied everywhere. This means that the resonance points can be chosen freely.

As mentioned above, the individual cylindrical decelerator stages consist of two types of approximate Halbach cylinders.  In an ideal case, the local magnetization is given by
\begin{equation}
\vec{M} = M_{r} \left[ \cos \left(K \phi \right) \hat{\textbf{\i}} + \sin \left( K \phi \right) \hat{\textbf{\j}} \right],
\label{Magnetization_Equation}
\end{equation}
where $M_{r}$ is the remanent magnetization amplitude, $\hat{\textbf{\i}}$ and $\hat{\textbf{\j}}$ form a local Cartesian basis perpendicular to the cylinder axis, $\phi$ is the polar angle, and $K$ is the number of rotations made by the local magnetization around a closed path that encompasses the inner aperture \footnote{This is equivalent to the alternative definition $\vec{M}=M_{r} \left[ \cos \left(m \phi \right) \hat{\rho} + \sin \left( m \phi \right) \hat{\phi} \right]$ for integer $m$, provided the substitution $K$=$m$+1 is made.}.  In general, choice of $K$ yields a ``2($K$-1)''-pole field, where the field magnitude in the bore depends on the radius as $| B | \sim r^{K-2}$.  We use $K$=2 to produce a region of strong uniform magnetic field, and $K$=6 to guide molecules in weak-field seeking states. For molecules to be pumped from the strong- to the weak-field seeking state, they must pass through regions of sufficiently small magnetic field that they can come into resonance with $L_{s \rightarrow w}$. If they repeatedly fail to do that, they will be lost from the decelerator. Therefore, we would like the guiding magnets to have a large area where the field is low, and steep potential walls that do the guiding. The $K$=6 cylinder has this property, which is why we choose it. We have found that these large weak-field regions are essential for efficient deceleration, as discussed further in Sec.~\ref{Trajectory_Simulations}.

\begin{figure}[!htb]
\centering
\includegraphics[width=\linewidth]{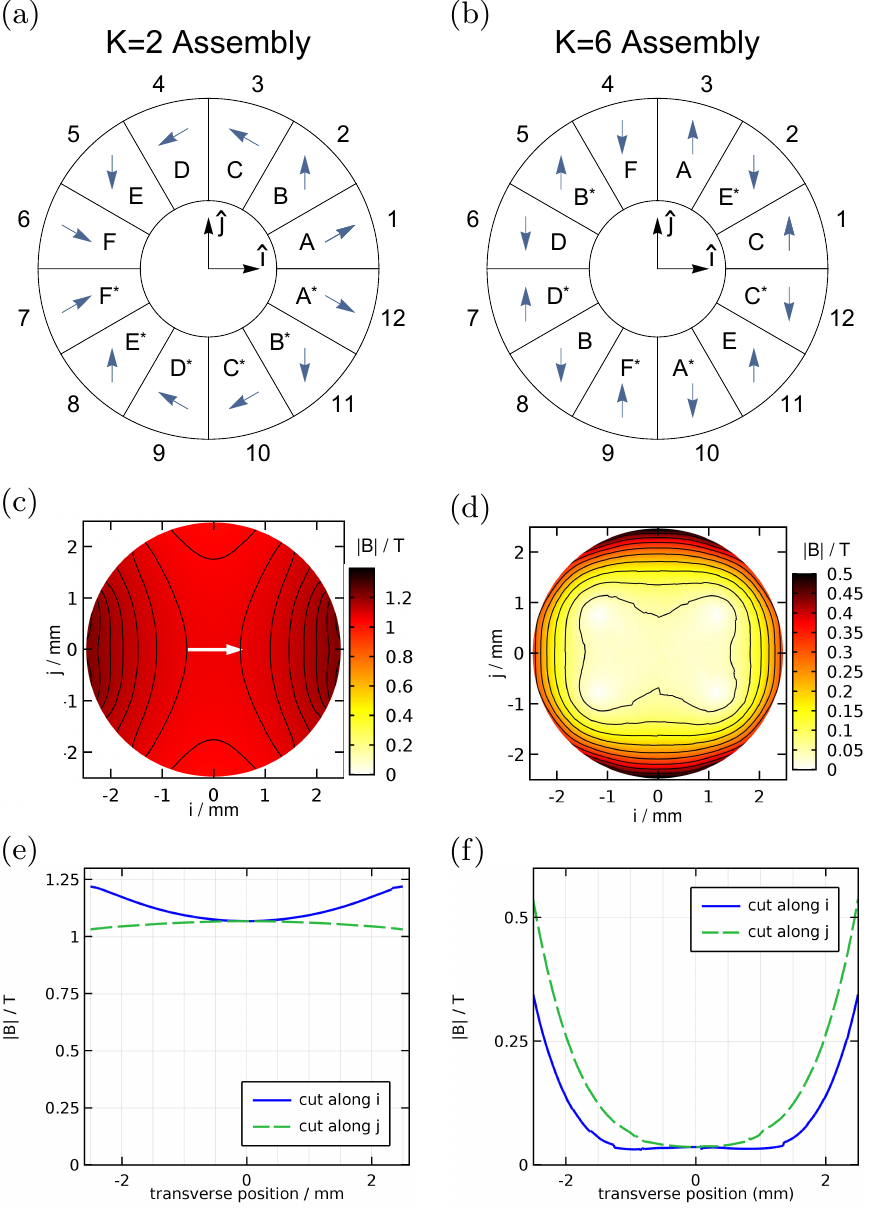}
\caption{$K$=2 (left column) and $K$=6 (right column) approximate Halbach cylinders. (a) and (b) show how each cylinder is constructed from twelve discrete pie-shaped wedge magnets using only six unique magnets, labeled A-F. (c) and (d) are plots of the magnetic field magnitudes through a central slice of each cylinder type, for the same geometric parameters used in the decelerator.  (e) and (f) show further cuts through these surfaces along the two principal axes.}
\label{Real_Magnets}
\end{figure}

In practice, it is difficult to manufacture strong permanent magnets with locally varying magnetization. Instead, we approximate each of the Halbach cylinders using 12 wedges, as shown in Figure~\ref{Real_Magnets}(a)-(b). The magnetization of each wedge relative to the coordinate axes can be expressed by equation~(\ref{Magnetization_Equation}) with the substitution $ \phi \rightarrow  \frac{2 \pi}{W} \left(w-\frac{1}{2} \right) $, where there are $W$ discrete wedges labeled by $w \in \{1,\ldots,W\}$.  Choosing $W$=12 and recognizing the symmetry of the wedge magnet array, one finds that only six unique magnets are required to construct either the $K$=2 or $K$=6 Halbach cylinders.  These are denoted A-F in Figure~\ref{Real_Magnets}(a)-(b), where a superscript $^{*}$ indicates a wedge has been flipped into the page. The required magnetization directions relative to the radius vector that bisects the wedge are 15$^{\circ}$,45$^{\circ}$,75$^{\circ}$,105$^{\circ}$,135$^{\circ}$, and 165$^{\circ}$ for A-F, respectively.  Figure~\ref{Real_Magnets}(c)-(f) shows the resulting magnetic fields as calculated by finite-element methods, with the left (right) column showing results for the $K$=2 ($K$=6) cases.  The geometry is identical to that of the final decelerator design, described above.  As shown, the $K$=2 and $K$=6 cylinders produce the desired field characteristics for strong-field and guiding-field cases, respectively.

\section{Application to CaF}
\label{CaF}
In the rest of this paper we explore the dynamics of molecules traveling through the Zeeman-Sisyphus decelerator, using calcium monofluoride (CaF) as a prototypical molecule. CaF is amenable to laser cooling~\cite{Zhelyazkova2014}, with at least two optical cycling transitions known, being $A^{2}\Pi_{1/2}$($v=0$,$J$=1/2)--$X^{2}\Sigma^{+}$($v$=0,$N$=1) and $B^{2}\Sigma^{+}$($v$=0,$N$=0)--$X^{2}\Sigma^{+}$($v$=0,$N$=1)~\cite{Tarbutt2015}. In strong magnetic fields, both the $X$ and $B$ states have large Zeeman shifts approximately equal to that of a free electron.  Conversely, the $A$ state has a a small magnetic moment because the spin and orbital magnetic moments are almost exactly equal and opposite. In the $X$ and $B$ states the electron spin is uncoupled from all other angular momenta in the large magnetic fields of the decelerator, and the Zeeman sub-levels are characterized by $m_{s}$, the projection of the spin onto the field axis. Since $m_{s}$ cannot change in an electric dipole transition, the optical pumping between strong- and weak-field seeking states cannot be achieved on the $B$--$X$ transition.  Because the spin-orbit interaction of the $A$ state is vastly larger than the Zeeman shift at all relevant fields, the Zeeman sub-levels are characterized by $m_{J}$, the projection of the total electronic angular momentum onto the magnetic field axis. These levels are of mixed $m_{s}$ character, and so the optical pumping works well. Because of these features of the $A$--$X$ transition, the simplified scheme illustrated in Figure~\ref{basicIdeaFig} is a good representation of decelerator operation for this molecule and transition.

\begin{figure*}[!htb]
\centering
\includegraphics[width=\linewidth]{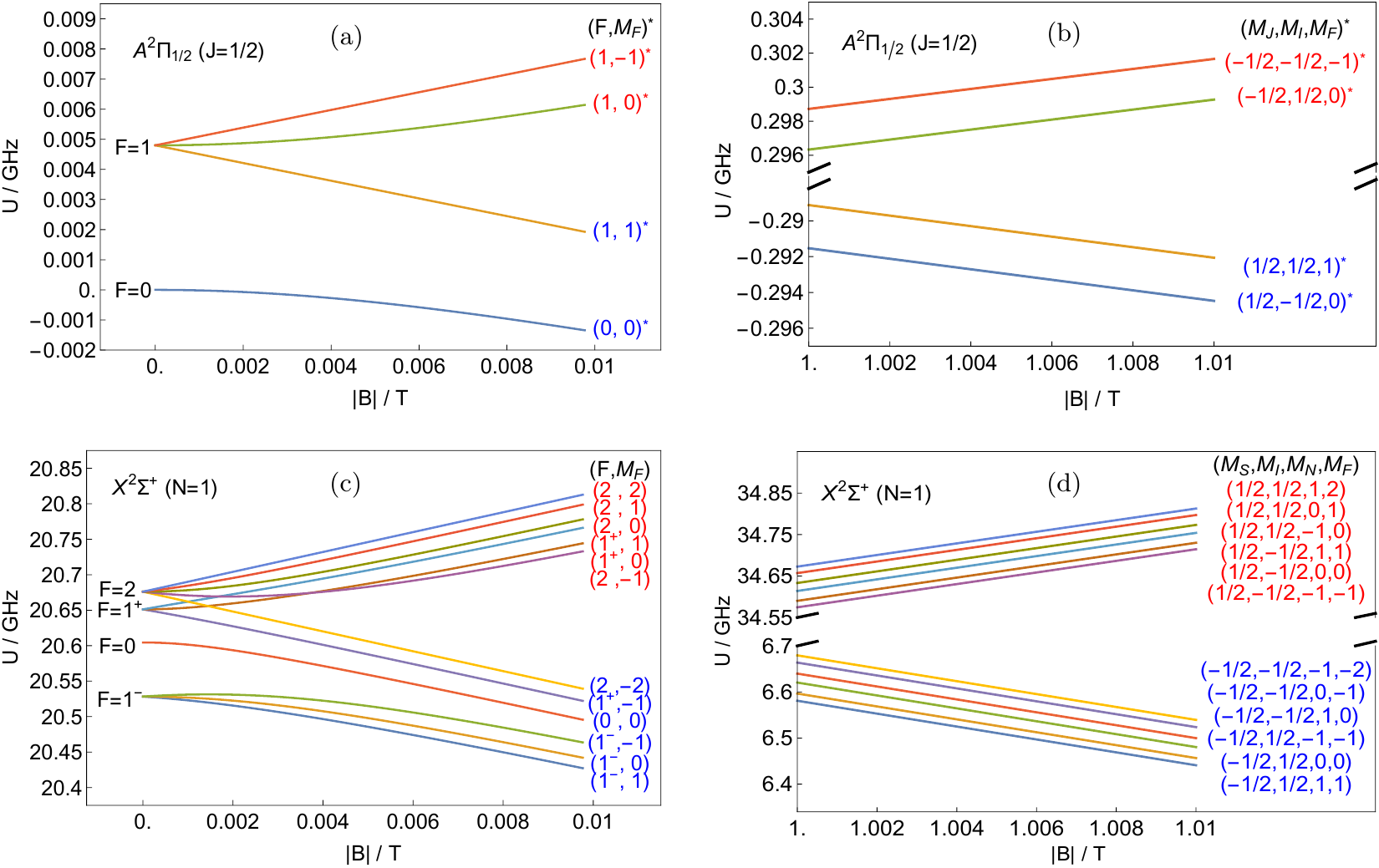}
\caption{Zeeman effect of the $X^{2}\Sigma^{+}$($v$=0,$N$=1) (lower plots) and the $A^{2}\Pi_{1/2}$($v$=0,$J$=1/2) (upper plots) levels of CaF in the low (left plots) and high (right plots) field limits. The ground state energy is defined relative to the $N$=0 level resulting in a constant offset of approximately 20.5~GHz. Quantum state labels are color coded according to whether they are wfs (red) or sfs (blue), consistent with Figure~\ref{basicIdeaFig}. The $*$ superscript identifies excited states. Note the different energy scales for the ground and excited states, as well as the broken energy axis in the high-field plots.}
\label{CaF_Zeeman}
\end{figure*}

Figure~\ref{CaF_Zeeman} shows the Zeeman shifts of the relevant states of CaF~\cite{Devlin2015}. The behavior of the ground (excited) state is shown in the lower (upper) plots, with the low (high) field regime shown on the left (right). Figure~\ref{CaF_Zeeman}(a) shows the Zeeman shifts in the $A$ state at low field. At zero field there are two hyperfine levels whose splitting is known to be smaller than 10~MHz. Following~\cite{Tarbutt2015}, we have set this splitting to 4.8~MHz, though the exact value is too small to be of any relevance. There are four magnetic sub-levels labelled by $(F,M_{F})$ in weak fields. In strong fields, they are labelled by $(M_{J},M_{I})$ and form a wfs and sfs manifold as shown in Figure~\ref{CaF_Zeeman}(b). The individual components of each manifold have equal gradients with magnetic field, and they are spaced by about 2~MHz. Figure~\ref{CaF_Zeeman}(c) shows the shifts of the $X$ state in low fields. This state consists of four hyperfine components labeled by their total angular momentum as $F$=$\{1^{-},0,1^{+},2\}$ in ascending energy, where the $\pm$ superscripts act only to distinguish between the two $F$=1 levels.  These hyperfine levels split into 12 magnetic sublevels, six weak-field seeking and six strong-field seeking, each labelled by $(F,M_{F})$.  These two manifolds play the part of the single wfs and sfs ground states in the simplified picture of Figure~\ref{basicIdeaFig}.  Figure~\ref{CaF_Zeeman}(d) shows how these states shift at high magnetic field. The six levels of each manifold have a nearly uniform spacing of about 20~MHz, and they have equal gradients with magnetic field which is about 50 times larger than that of the $A$ levels. In this high-field regime, the states are properly labelled by $(M_{S}, M_{I}, M_{N})$. However, we choose to label each level at all fields according to the $(F,M_{F})$ state it becomes as the field is adiabatically reduced to zero.

We see from Figure~\ref{CaF_Zeeman} that the pump lasers must address transitions between multiple levels. Since the Zeeman shift is far larger than the splitting between the levels of both the wfs and sfs manifolds, the longitudinally varying magnetic field will bring the various transitions into resonance at slightly different longitudinal positions. This means that, despite the multiple levels, only one laser frequency is needed to optically pump molecules in one direction. However, the presence of multiple levels is expected to make the optical pumping more likely to fail.  Consider, for example, a ground state molecule in the $(2,-1)$ state entering a region of large magnetic field.  As the lowest level in the wfs manifold, this molecule will come into resonance with the pump laser at the most advanced position.  The excited state may decay to a different sublevel of the wfs manifold, and the difference in energy between these two sublevels might be large enough that the molecule is now too far out of resonance with the pump laser to be excited a second time.  This effect can be worsened by the non-zero Zeeman shift of the excited state sublevels.  Specifically, the new ground sublevel may predominantly couple to the opposite Zeeman manifold in the upper state compared to the initial ground level, taking the molecule even further from resonance.  Because of these multi-level effects, a molecule may pass through the resonance point without being optically pumped and will continue through a stage of the decelerator in the wrong state.  All is not lost, however, because a molecule that fails to be pumped has a second chance on the opposite side of the potential energy hill.  These effects are not captured by the simple model presented in Sec.~\ref{Zeeman_Sisyphus_Deceleration}, but are included in the simulations discussed in Sec.~\ref{Trajectory_Simulations}. The effects can be mitigated by reducing the magnetic field gradient, increasing the laser power, or adding sidebands to the laser to increase its frequency spread.

\begin{figure*}[!htb]
\centering
\includegraphics[width=\linewidth]{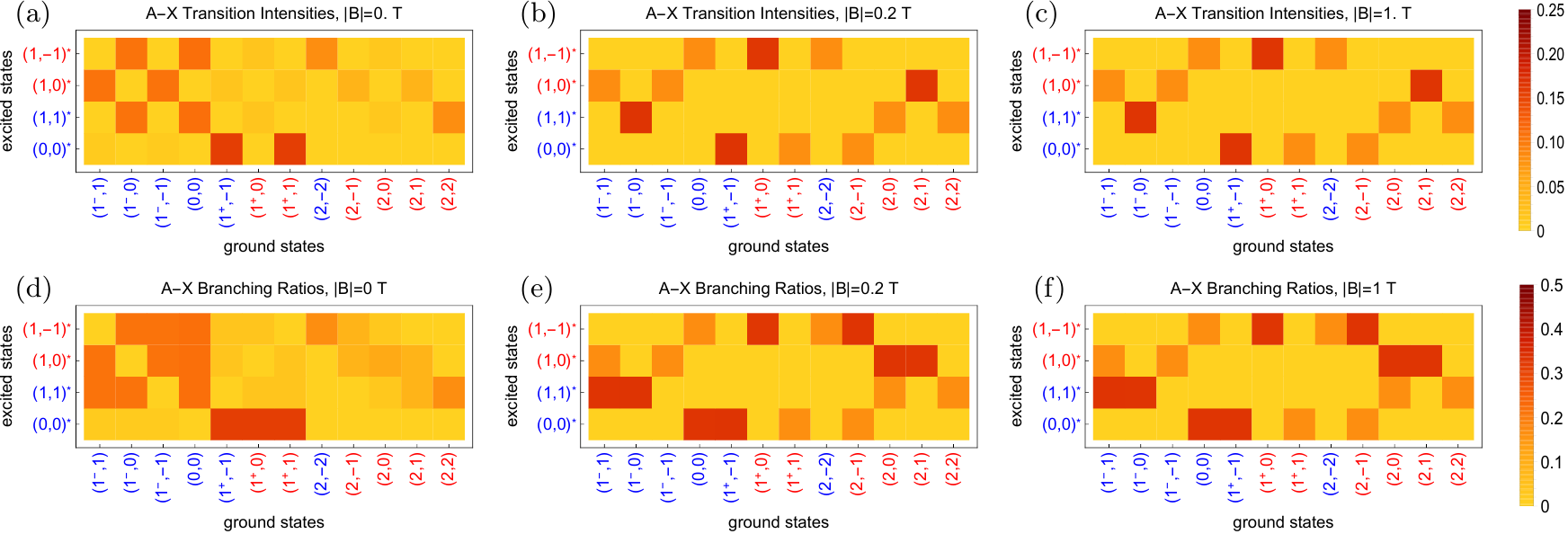}
\caption{Transition intensities (top row) and branching ratios (bottom row) between the $X^{2}\Sigma^{+}$($v$=0,$N$=1) and $A^{2}\Pi_{1/2}$($v$=0,$J$=1/2) states, for magnetic field strengths of 0 (left column), 0.2 (middle column), and 1~T (right column).  In calculating the transition intensities we assume pump light linearly polarized perpendicular to the magnetic field axis.  Numerical values for both the transition intensities and branching ratios are presented in Section~\ref{Appendix}, together with a discussion of the effects of the variations in pointing of $\vec{B}$ at the optical pumping locations.}
\label{Excite_And_Decay_Strengths}
\end{figure*}

A second potential problem for the optical pumping is level crossings with other quantum states not yet considered. For example, the sfs manifold of the $N$=1 ground state crosses the wfs manifold of the $N$=0 state at a field of $\approx$0.75~T. A molecule transferred to $N$=0 at this crossing will be lost from the decelerator since the lasers are tuned to drive the cycling transition from $N$=1 and do not address the $N$=0 levels. Fortunately, there is no coupling between these two states because they are of opposite parity and the magnetic field can only couple states of the same parity. An electric field turns the crossing into an avoided crossing and so must be kept sufficiently small. The electric field arising from the motion of the molecules through the magnetic field is too small to be of concern. The situation is similar near 1.5~T, where the wfs $N$=1 manifold crosses the sfs $N$=2 manifold. The first problematic crossing is between $N$=1 and $N$=3, since they have the same parity, but this occurs near 2.5~T, which is well above the fields present in the decelerator.

A third concern for the reliability of the optical pumping is that other transitions from the $N$=1 state might come into resonance with the laser light and transfer molecules out of the cycling transition. In this case, the only such transition is the $Q$(1) transition, which is approximately 30~GHz higher in frequency than the $P$(1) cycling transition at zero field. For a typical choice of detuning, the $Q$(1) transition comes into resonance with $L_{s \rightarrow w}$ when the field is about 1.75~T. Fortunately, this is higher than the largest field present in the decelerator. We see that, at least for CaF, no other states or transitions play any role in the decelerator and our analysis can focus solely on the 12 ground states and 4 excited states shown in Figure~\ref{CaF_Zeeman}. This good fortune does not necessarily carry over to other molecules of interest; a similar analysis should be completed for each case.

To understand the optical pumping of the multi-level CaF system in the decelerator, we have calculated the relative transition strengths between each of the ground and excited sub-levels for various magnetic field strengths and laser polarizations. Figure~\ref{Excite_And_Decay_Strengths} shows the transition intensities for excitation out of the ground states (top row) and the branching ratios for the decay of the excited states (bottom row), for magnetic fields of 0 (left column), 0.2 (middle column), and 1~T (right column).  The last two field values are typical values where the two optical pumping processes occur.  In calculating the transition intensities we have taken light linearly polarized perpendicular to the strong magnetic-field direction, which is the configuration used in the decelerator. We see that the transition intensities and branching ratios change significantly between the zero- and nonzero-field cases, but change very little between the two non-zero field values.  In fact, we find that all branching ratios change by less than 2\% in absolute value as the field increases beyond 0.03~T. This makes sense in the context of Figure~\ref{CaF_Zeeman}(a,c) where we can see (by extrapolation) that the levels are already grouped into well-spaced wfs and sfs manifolds once the field reaches this value. Since the optical pumping occurs at fields much higher than this, we take the branching ratios and transition intensities to be constants in the numerical simulations presented in Sec.~\ref{Trajectory_Simulations}.  This choice is discussed further in the Appendix.

Let us consider in more detail a particular optical pumping event. As our example, we consider a molecule in the sfs state $(1^{-},-1)$ entering a region of weak magnetic field and coming into resonance with the $L_{s \rightarrow w}$ laser. The laser drives, almost exclusively, the transition to the $(1,0)^{*}$ excited state (see Figure~\ref{Excite_And_Decay_Strengths}(b), 3rd column). This state can decay to the wfs states $(2,0)$ or $(2,1)$, each with 33\% probability (see Figure~\ref{Excite_And_Decay_Strengths}(e), 2nd row) and so there is a 66\% probability that the molecule switches between the sfs and wfs manifolds after scattering a single photon.  The excited state can also decay to the sfs states $(1^{-},1)$ or $(1^{-},-1)$, each with 17\% probability.  Both states remain near resonance with the pump laser, and so the molecule is likely to be re-excited, again to the $(1,0)^{*}$ state, giving it a second 66\% chance of switching between sfs and wfs manifolds.  In the notation of equation~\ref{chi}, $p$=2/3, and a value of $\bar{n}$=4 is sufficient to ensure $\chi>$0.98.  After successful optical pumping, the molecule has a 50\% chance of being in either of the two participating wfs ground states.

As an example of a state that does not optically pump as efficiently, consider a molecule in the sfs $(1^{-},0)$ state under the same conditions.  Again, the pumping laser almost exclusively drives a single transition, in this case to $(1,1)^{*}$.  The subsequent spontaneous decay takes the molecule to a wfs state [either $(2,0)$ or $(2,2)$] only 33\% of the time, giving $p$=1/3. The molecule is returned to an sfs state [either $(1^{-},1)$ or the original $(1^{-},0)$ state] with 66\% probability. The decay to $(1^{-},1)$ is particularly troublesome, because this state couples only to $(1,0)^{*}$ in the wfs upper manifold, whereas the original $(1^{-},0)$ state couples only to $(1,1)^{*}$ in the sfs upper manifold.  Thus, the resonance condition may be lost due to the Zeeman shift of the excited state.  This is a greater concern for pumping from the wfs to the sfs ground-state manifolds, since that process occurs at larger fields where the upper-state manifolds are further separated.

\begin{figure}[!htb]
\centering
\includegraphics[width=0.9\linewidth]{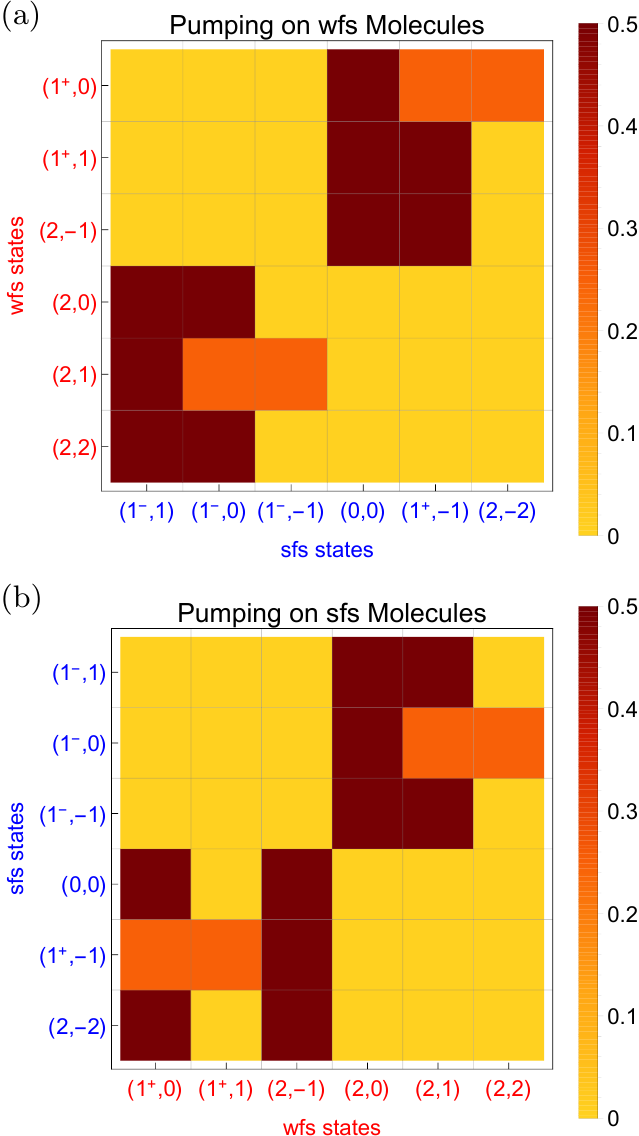}
\caption{Transformations of hyperfine state populations due to optical pumping with light linearly polarized along $\hat{y}$, assuming perfect transfer between the wfs and sfs manifolds.  States being pumped appear on the vertical axis. The horizontal axis gives the relative population in each of the ground states following optical pumping.}
\label{CaF_Pumping}
\end{figure}

Repeating the optical-pumping analysis for each of the twelve ground states reveals that eight of the states have $p$=2/3, while the remaining four have $p$=1/3.  These four all exhibit the behavior described above where a failure to optically pump may take the molecule to a state where the optical pumping transition is further from resonance due to the excited-state Zeeman splitting. None of the eight states that pump with high efficiency exhibit this behavior.

Assuming that the optical pumping proceeds with unit probability despite the aforementioned difficulties, the molecular ensemble continually exchanges population between the wfs and sfs manifolds at each resonance point.  Details of the population transfer between the two ground state manifolds is summarized in Figure~\ref{CaF_Pumping}, which can be derived by continually propagating the set of ground states through the excitation and decay processes presented in Figure~\ref{Excite_And_Decay_Strengths}.  Here, populations pump from initial states indicated by column to final states indicated by row.  The eight states that pump efficiently transfer strongly to only two states in the opposite manifold, while the four that pump less efficiently are transferred to three states in the opposite manifold.

\section{Trajectory Simulations}
\label{Trajectory_Simulations}
We now study the dynamics of CaF molecules in the decelerator in more detail by using trajectory simulations. A simulation takes as its input an initial phase-space distribution, a map of the magnetic field calculated using a finite element model, and a table of transition strengths and branching ratios between the ground and excited states, which we take to be independent of $|\vec{B}|$ as discussed in Sec.~\ref{CaF}. The direction of the magnetic field changes little over the set of positions where the optical pumping occurs, being purely $\pm \hat{x}$ to a good approximation.  For all the simulations presented here the pump lasers are linearly polarized along $\hat{y}$ and the transition strengths are independent of whether $\vec{B}$ is parallel or anti-parallel to $\hat{x}$.  The effects of variations in the pointing of $\vec{B}$ at the optical pumping locations is discussed further in Section~\ref{Appendix}. In all our simulations, the laser intensity profile is assumed to be Gaussian with a full width at half maximum (FWHM) of 5~mm, the same as the inner diameter of the decelerator.  We use linear Zeeman shifts and the hyperfine splittings shown in Figure \ref{CaF_Zeeman}(b),(d). This assumes that molecules never experience magnetic fields below about 0.01~T, which is a good assumption for nearly all trajectories.  This assumption is discussed further in Section~\ref{Appendix}.  The cycling transition for this system, consisting of the 12 ground states and 4 excited states, is considered to be closed; the excited states always decay to one of the 12 ground states.  In reality, some repumping of population that leaks into $v$=1 may be required.

The simulation propagates each molecule through the decelerator under the action of the force $\vec{F} = - \nabla U(|\vec{B}|)$, and keeps track of its state as it is optically pumped. During a time step $\Delta t$, the probability of a molecule initially in ground state $i$ scattering a photon via excitation to state $j$ is calculated as
\begin{equation}
P_{j} = R \, T_{ij} \,\Delta t.
\end{equation}
Here, $R$ is given by equation~(\ref{scattering_rate}) and $T_{ij}$ is the pre-calculated transition intensity between states $i$ and $j$. The total probability of scattering a photon during this time step is $P = \sum_{j} P_{j}$. We choose the time step so that $P \ll$1, typically $\Delta t$=10~ns. A random number, $r$, is selected from a uniform distribution between 0 and 1. If $r>P$ no transition occurs. If $r<P$ a transition occurs and the excited state is selected at random according to the relative probabilities $P_{j}$. The molecule then decays, with the final ground state selected randomly according to the pre-calculated branching ratios. The photon is emitted in a random direction chosen from an isotropic distribution. The new position and speed at the end of the time step are then calculated, including the small changes in momentum due to the absorbed and spontaneously emitted photons. The simulation then proceeds to the next time step.

The initial phase-space distribution used for the simulations starts all molecules at $t$=0, $z$=0, but with a range of initial forward speeds.  For the transverse degrees of freedom,
we typically use a distribution that is uniform in the range from $\pm$2.5~mm and $\pm$7.5~m/s for both transverse dimensions. This range is larger than the decelerator can accept, and so most molecules are lost via collisions with the inner magnet surfaces in the first $\sim 25$~cm of the decelerator.  By overfilling the transverse phase space in this way, we ensure that the molecular distributions at the exit of the decelerator are indicative of the deceleration dynamics and not the particular choice of initial conditions.  Combined, the initial transverse and longitudinal phase-space extents of the molecular distribution do an acceptable job of simulating molecules with not only differing forward speeds but also differently directed initial velocity vectors.

\subsection{Guiding Performance}
\label{Guiding_Performance}
We first turn off the optical pumping light and study the performance of the magnet array as a guide for molecules in wfs states.  This is useful in identifying dynamical instabilities that arise from the coupling of longitudinal and transverse motions, and helps to elucidate why molecules are lost as they are decelerated~\cite{Meerakker2006}.

\begin{figure}[!htb]
\centering
\includegraphics[width=\linewidth]{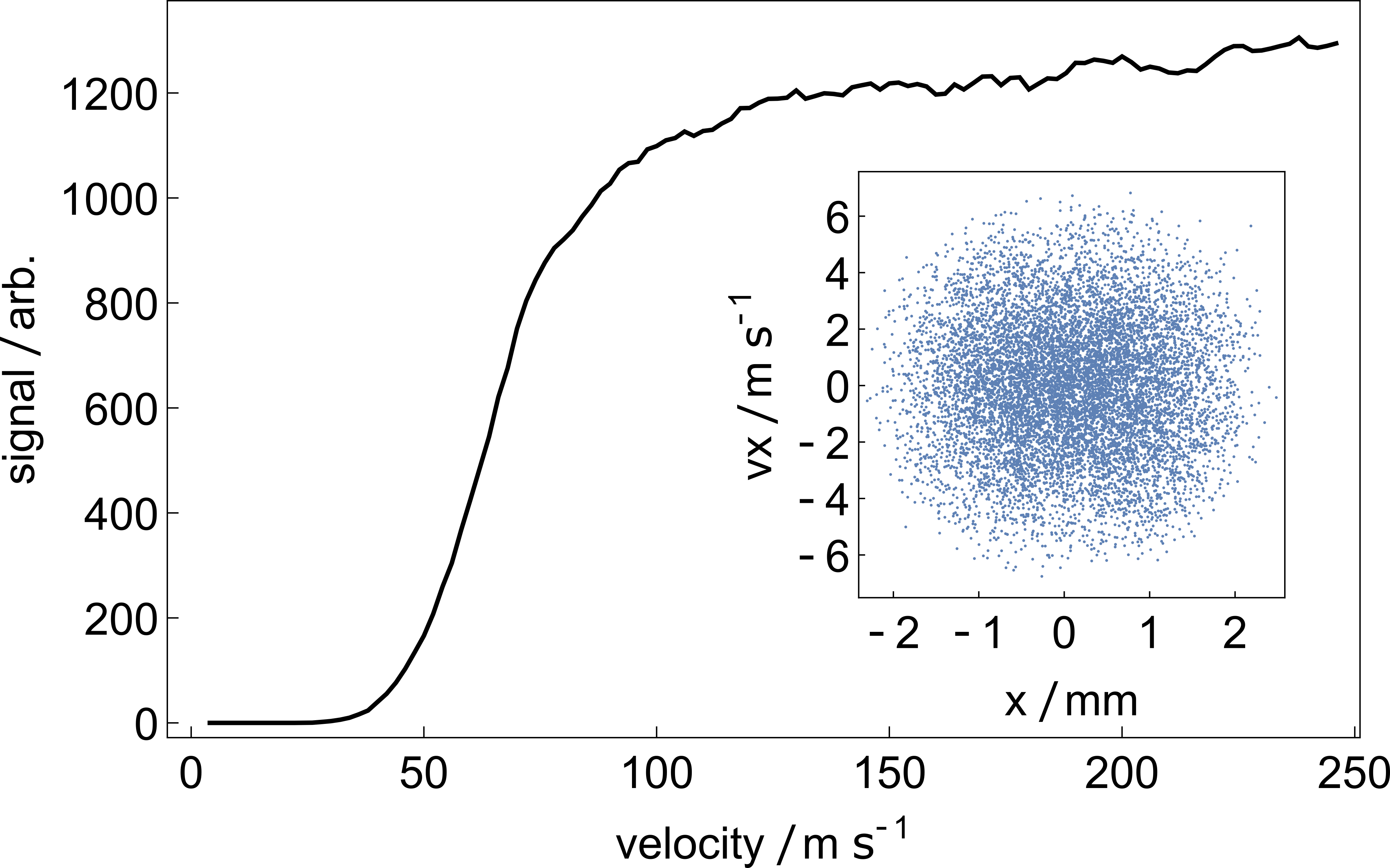}
\caption{Relative number of molecules transmitted to the end of a 1~m long decelerator used as a guide for weak-field seeking molecules, as a function of their forward speed. The inset shows the transverse phase-space distribution of molecules that exit the decelerator.}
\label{Guiding_Figure}
\end{figure}

Figure~\ref{Guiding_Figure} shows the relative number of molecules reaching the end of the magnet array as a function of their initial forward speed.  Molecules with $v_{z}\le$14~m/s have insufficient kinetic energy to climb over a single potential energy hill, so this sets a lower limit to the speed a molecule can have in order to reach the exit. Above 100~m/s, the number guided does not depend strongly on the speed, but below 100~m/s the number that reach the end falls off rapidly with decreasing speed. This is because there are stronger guiding forces in the guiding magnets than in the strong-field magnets. The slow molecules are guided too strongly by the guiding magnets and are then lost in the strong-field magnets where the guiding is weak. Moreover, the modulation of the transverse guiding can couple energy from the longitudinal motion into the transverse motion, causing further loss. These effects set in once $v_{z}\le$4$L/T$, where $L$=2~cm is the spatial periodicity of the magnet array and $T$ is the transverse oscillation period. The guide is not harmonic, so there is a range of oscillation periods, but $T \approx$1~ms is typical. Thus, we expect the losses to set in when $v_{z}\approx$80~m/s, which is roughly what we observe in the simulations.  Because of this loss mechanism, it is advantageous to decelerate the molecules as rapidly as possible once they reach low speed. Fortuitously, the highest optical-pumping efficiency, and therefore the highest deceleration, is naturally realized by the slowest molecules.  We note that the low-speed stability can be improved by reducing the spatial periodicity ($L$) near the end of the decelerator, or increasing the bore size of the magnets near the end so that the oscillation period ($T$) increases.

The inset to Figure~\ref{Guiding_Figure} shows the transverse phase-space distribution of molecules that exit the guide. The spatial extent of $\pm$2.5~mm is set by the bore diameter of the magnets, and the velocity spread of $\pm$6~m/s is set by the energetic depth of the guide.

\begin{figure*}[!htb]
\centering
\includegraphics[width=\linewidth]{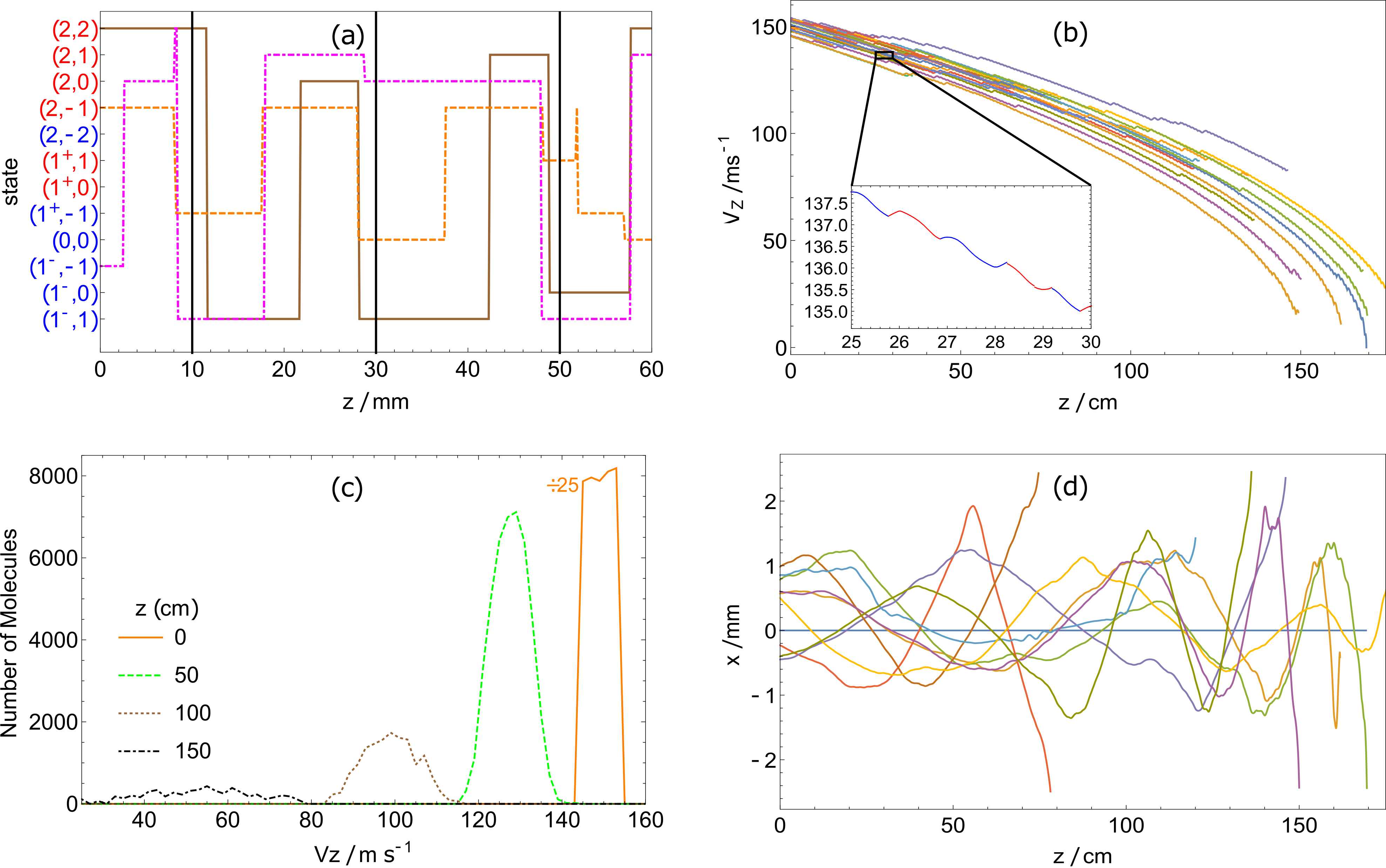}
\caption{Trajectory simulations for an initial distribution of molecules with speeds centered around 150~m/s.  Pump laser powers were set to 200~mW with $\Delta_{w \rightarrow s}=13.75$~GHz, $\Delta_{s \rightarrow w}=2.5$~GHz. (a) Tracking the state of three sample molecules as they propagate through the first few stages of the decelerator. Solid vertical lines are the positions near which wfs states (labeled red) should be optically pumped to sfs states (labeled blue). (b) Forward velocity of a group of molecules as they propagate through the decelerator.  The inset shows a typical trajectory over 5~cm. (c) The forward-velocity distribution at various longitudinal positions. Note that the initial distribution is 25 times larger than shown. (d)  Trajectories in the $x z$-plane.}
\label{Simulations_Trajectories}
\end{figure*}

\subsection{Deceleration Performance}
\label{Deceleration_Performance}
Figure~\ref{Simulations_Trajectories} follows some molecules, all with initial velocities around 150~m/s, as they propagate through the decelerator.  Here, the pump laser powers are 200~mW with detunings of $\Delta_{w \rightarrow s}=13.75$~GHz and $\Delta_{s \rightarrow w}=2.5$~GHz.  In Figure~\ref{Simulations_Trajectories}(a), quantum-state tracking for three molecules is shown for the first few cm of the decelerator, with the solid vertical lines representing strong-field regions near which molecules should optically pump from wfs states (red) to sfs states (blue).  Transitions from sfs to wfs states should occur in the weak-field regions between the vertical solid lines.  In most cases, the optical pumping is successful. At $z\simeq28$~mm there is an example of a failure to switch. The molecule in the wfs state $(2,1)$ is excited at this position but decays to the $(2,0)$ state which is another wfs state. The molecule is not excited a second time, and so travels through a stage of the decelerator in the wrong state. Another interesting example occurs near $z=50$~mm. The molecule in the wfs state $(2,-1)$ first switches to the wfs state $(1^{+},1)$, then back to $(2,-1)$ before finally being pumped to the sfs state $(1^{+},-1)$.  Figure~\ref{Simulations_Trajectories}(b) shows the molecules approximately following the $v-z$ curves expected for a constant deceleration. Occasional failures to optically pump can be seen as horizontal propagation with no net change in forward velocity. The inset follows a single molecule over a short region of the decelerator, showing five switches between wfs and sfs states and the associated deceleration. Figure~\ref{Simulations_Trajectories}(c) shows the distribution of forward velocities at various longitudinal positions with the initial distribution divided down for easier comparison.  The velocity distribution spreads as molecules are decelerated, in accordance with the dynamics of constant deceleration from various initial velocities, as discussed in Sec.~\ref{Zeeman_Sisyphus_Deceleration}. Approximately 10\% of the population present at 50~cm, with a mean speed of 130~m/s, successfully propagates to 150~cm where the mean speed has been reduced to 55~m/s. Figure~\ref{Simulations_Trajectories}(d) shows some molecular trajectories in the $x z$-plane. We see that some molecules transversely oscillate through the decelerator on stable trajectories, showing the effect of the transverse guiding. Some hit the walls at $x$=$\pm $2.5~mm before reaching the end of the decelerator, while others come to rest before they get to the decelerator's exit.  Both cases are indicated by a trajectory abruptly ending. From the simulation, we find that the average number of photons scattered by molecules that reach $z$=150~cm is 225, corresponding to 1.55 photons per resonance point. This is near the expected value for $p$=2/3, indicating that the few states which pump less efficiently do not play a large role in the deceleration dynamics.

\begin{figure}[!htb]
\centering
\includegraphics[width=\linewidth]{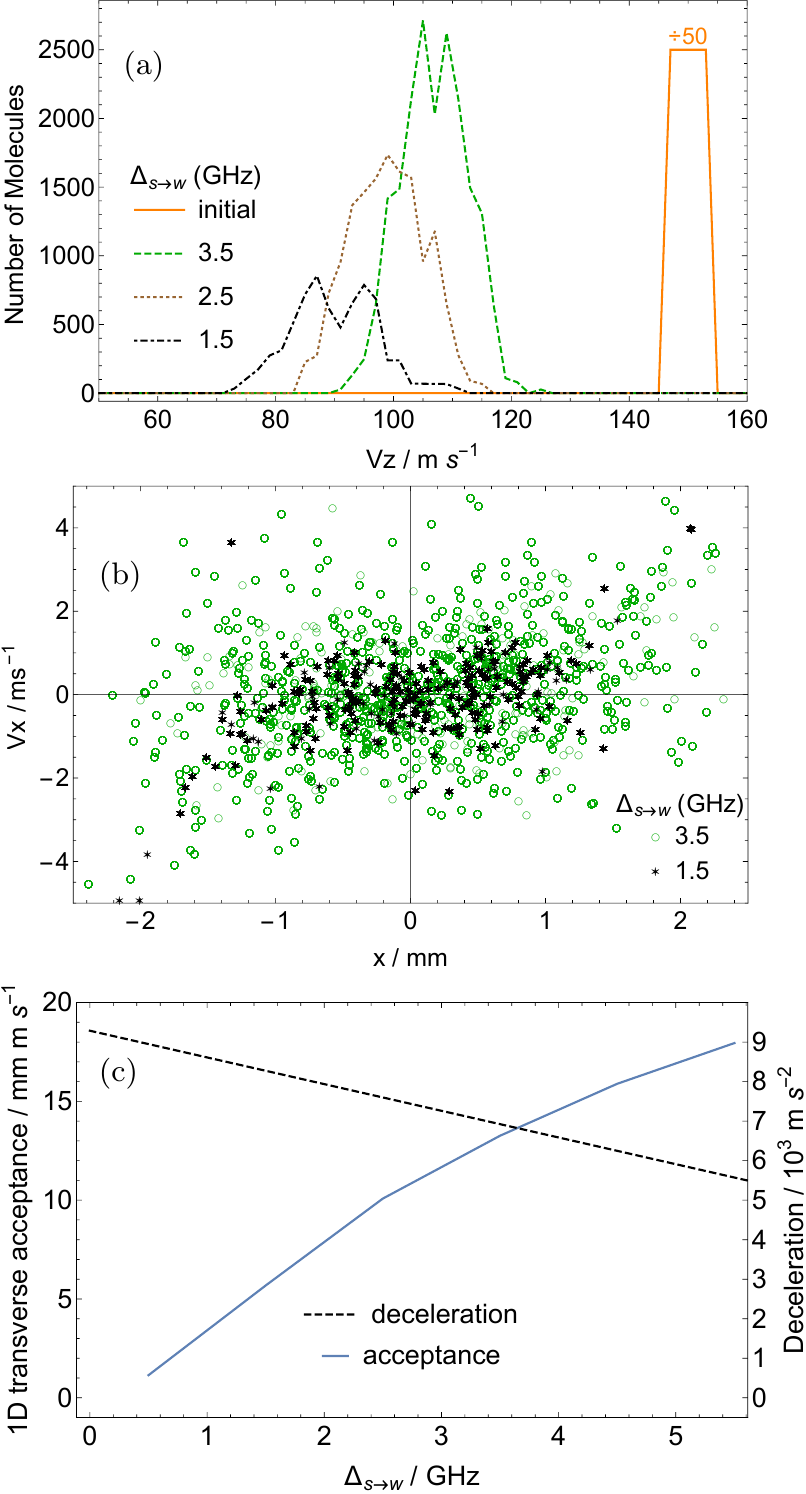}
\caption{Exploring the effect of $\Delta_{s\rightarrow w}$.  (a)  Longitudinal velocity distributions at $z=$1~m.  Smaller $\Delta_{s\rightarrow w}$ gives more deceleration but fewer molecules.  (b)  Final transverse phase-space distribution of those that reach $z$=1~m. (c) Deceleration and transverse acceptance as a function of $\Delta_{s\rightarrow w}$.}
\label{Simulations_Distributions}
\end{figure}

As discussed in Sec.~\ref{Zeeman_Sisyphus_Deceleration}, setting $\Delta_{s \rightarrow w}$ close to zero gives the maximum deceleration but reduces the transverse stability because molecules in sfs states may never reach low enough fields to come into resonance with $L_{s \rightarrow w}$. Figure~\ref{Simulations_Distributions} explores this effect. Here, we use the same simulation settings as before, except that the decelerator length is fixed at 1~m and $\Delta_{s \rightarrow w}$ is varied. Figure~\ref{Simulations_Distributions}(a) shows how $\Delta_{s \rightarrow w}$ influences the final velocity distribution. As expected, bringing $L_{s \rightarrow w}$ closer to zero reduces the final average speed, but also reduces the number of molecules at the exit of the decelerator. Figure~\ref{Simulations_Distributions}(b) shows the final phase-space distribution, in one transverse dimension, of those molecules that successfully exit the decelerator, for two different values of $\Delta_{s \rightarrow w}$. In decelerator terminology, the set of stable molecule positions and velocities defines the transverse phase-space acceptance, though in this case the concept is less well defined because the stochastic nature of the optical pumping means that a molecule can be lost even though it appears to be well inside the acceptance region.

To understand in more detail why the acceptance decreases with decreasing $\Delta_{s \rightarrow w}$, consider the magnetic fields experienced by molecules at various distances from the decelerator axis.  The on-axis field varies between 0 and 1~T, but further away from the axis the field does not reach low values. There will be some radius where the minimum field is above that required to bring molecules into resonance with $L_{s \rightarrow w}$ at all longitudinal positions.  Beyond this radius, optical pumping out of the sfs states fails, and the molecules stuck in sfs states are anti-guided and lost. For our magnet geometry and $\Delta_{s \rightarrow w}$=3.5~GHz, optical pumping should cease beyond a radius of 1~mm, consistent with the observed region of transverse stability shown. Figure~\ref{Simulations_Distributions}(c) plots the deceleration and the phase-space acceptance in one transverse direction, both as functions of $\Delta_{s \rightarrow w}$. The deceleration is determined using Equation~\ref{deceleration}. The acceptance is determined in an approximate way by calculating the area of an ellipse that encloses 90\% of the transverse phase-space distribution of molecules exiting the decelerator. We see that there is a modest reduction in deceleration as $\Delta_{s \rightarrow w}$ is increased from 0.5 to 5~GHz, but a very large increase in the transverse phase-space acceptance. Unless there is a strong penalty for having a longer decelerator, it is best to keep $\Delta_{s \rightarrow w}$ relatively large to give the largest decelerated flux.

\subsection{Prospects for MOT loading and comparison with other deceleration methods}
\label{MOT_Prospects}
An attractive application of Zeeman-Sisyphus deceleration is the production of slow molecules for loading into a magneto-optical trap (MOT).  We can use our trajectory simulations to estimate how many molecules might be loaded using this technique, and compare the result to other deceleration methods.  There are many available deceleration techniques, and they depend strongly on the choice of molecule and molecule source. We limit our discussion to deceleration of CaF molecules in the $N$=1 state, emitted by a cryogenic buffer-gas source. Unless otherwise noted, we assume a CaF beam with 10$^{11}$~molecules per steradian per shot in $N$=1, and a distribution of forward speeds approximated by a Gaussian distribution with a 150~m/s mean and a FWHM of 93~m/s~\cite{Truppe2016}.  The distance from source to detector will be fixed at 1.3~m. In all comparisons, we assume a 10~cm free-flight distance between the source and the location where deceleration begins, as is typically needed due to geometric or pressure constraints. The detector area is taken to be a circle of diameter 5~mm, which is equal to the diameter of the decelerator aperture described in this paper.

With the Zeeman-Sisyphus decelerator design presented above, molecules with speeds below 14~m/s cannot get over the final potential hill and are either lost transversely or trapped by the magnetic field.  This could be mitigated by using a weaker magnetic field near the decelerator exit.  Instead, we take as our figure of merit the number of molecules exiting the decelerator below 50~m/s, since this is approaching the MOT capture velocity and slow enough that a short distance of direct laser slowing could then be used to load the MOT.

\begin{figure}[!htb]
\centering
\includegraphics[width=\linewidth]{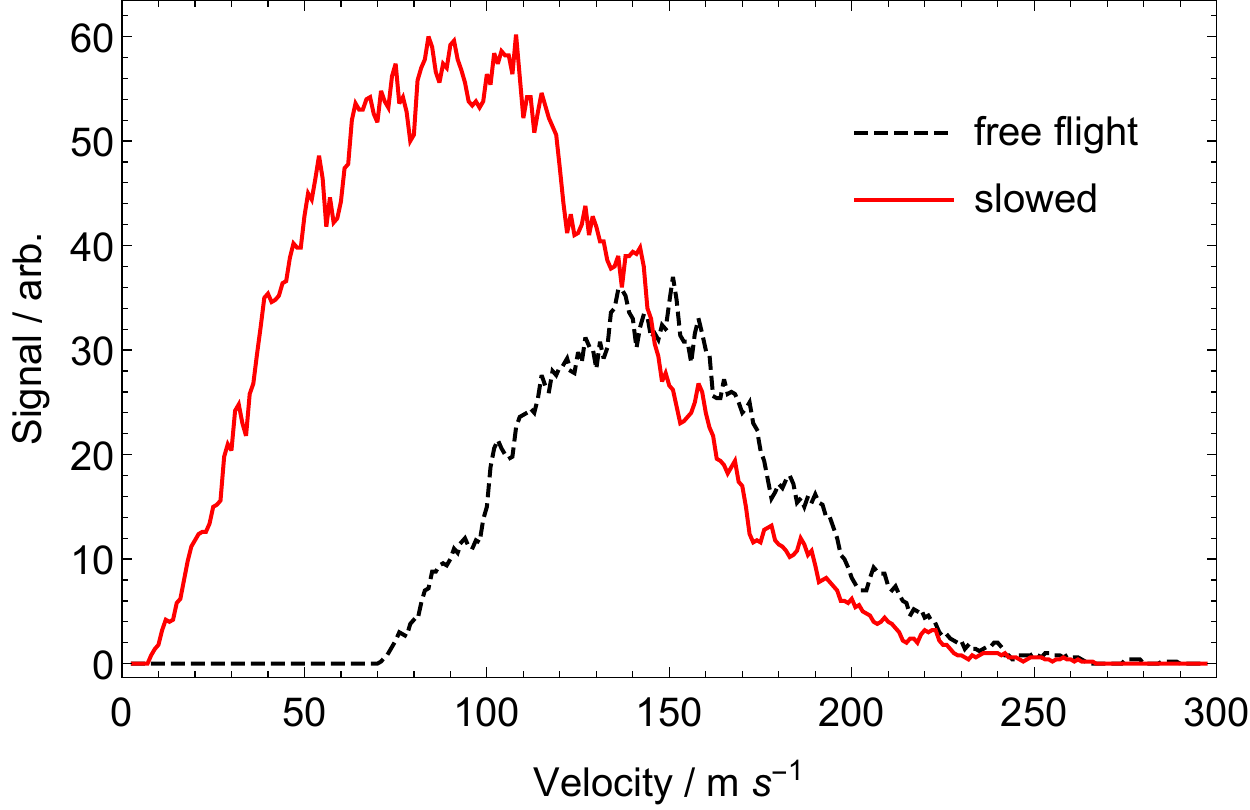}
\caption{Simulated velocity distributions of CaF molecules passing through the detection area at $z=1.3$~m, for free-flight (black, dashed) and following a 1.2~m Zeeman-Sisyphus decelerator (red, solid). The initial parameters are those of a buffer-gas beam of CaF as described in the text.}
\label{Slowing_Real_Beam}
\end{figure}

Figure~\ref{Slowing_Real_Beam} compares the simulated velocity distribution exiting a 1.2~m long Zeeman-Sisyphus decelerator, to the simulated distribution detected without the decelerator present. The decelerator parameters are identical to those described in Section~\ref{Deceleration_Performance} except the optical pumping laser detunings have been set to \{$\Delta_{w \rightarrow s}$, $\Delta_{s \rightarrow w}$\} = \{-14.25, 3\}~GHz.  This choice ensures molecules are optically pumped right at the potential-energy hilltop for wfs states.  As expected, the initially wide input velocity distribution from the buffer-gas source is both shifted down in velocity and broadened by the presence of the decelerator.  The effects of net guiding are also apparent; despite being slowed down, more molecules reach the detector when the beam is decelerated.  Also as expected, the distribution approaches zero at the velocity equivalent to the potential-energy hill height ($\sim$14~m/s), though there are a large number of molecules with velocities just above this limit.  In free flight, approximately 10$^{6}$ molecules per pulse pass through the detection  area. When the beam is decelerated, this number increases by a factor of 2.1, and 15\% of these have forward speeds below 50~m/s.  Thus, the decelerator produces about 3$\times$10$^{5}$ molecules in our chosen velocity range.

Let us compare this to the direct laser slowing results presented in~\cite{Truppe2016}, where the beam source and distance to the detector are the same. In this comparison the free-flight curves presented in~\cite{Truppe2016} are identical to Figure~\ref{Slowing_Real_Beam}, and represent the same absolute number of detected molecules.  In these experiments, about 2$\times$10$^{5}$ molecules are slowed to speeds below 50~m/s.  This is similar to the result above, indicating that Zeeman-Sisyphus deceleration is competitive with state-of-the-art direct laser slowing techniques.  The number of photons that have to be scattered is about $\sim$10$^{4}$ for direct laser slowing but only about 300 for Zeeman-Sisyphus deceleration.  This makes the decelerator a particularly attractive option for decelerating molecules where direct laser slowing may be impractical because the branching ratios are less favorable than for CaF.

Both the traditional~\cite{Bethlem1999} and traveling-wave~\cite{Osterwalder2010} Stark deceleration methods are also capable of slowing molecules into the velocity range of interest when starting with our beam parameters and using the same deceleration distance.  These methods are typically {\it not} well suited for deceleration of buffer-gas-cooled molecular beams due to the typically long ($
$1-10~ms) molecular pulses.  Our source is unusual because it produces a particularly short pulse, approximately 250~$\mu$s FWHM, making these time-dependent deceleration methods feasible.  To estimate the number of slow molecules that could be produced, we determine how many molecules from the initial distribution are within the longitudinal phase-space acceptance of the decelerator when it is turned on.  For the $N$=1 state of CaF, the maximum electric field that can be applied is approximately 30~kV/cm, which is a limitation in the traditional decelerator geometry. It is not a limitation for the traveling-wave decelerator, which by design uses smaller peak electric fields.  An acceleration of -8.3$\times$10$^{3}$~m/s$^{2}$ is sufficient to decelerate molecules from 150~m/s to 50~m/s in 1.2~m, corresponding to a synchronous molecule phase angle of 24.5$^{\circ}$ for the Stark decelerator. We calculate longitudinal phase-space acceptances of 65 and 16~mm$\times$m/s for the traditional and traveling-wave decelerator respectively.  The $\sim 3$ times larger solid angle subtended by the traveling-wave decelerator makes up most of the difference, and so both methods yield roughly the same number of slow molecules, approximately 3$\times$10$^{5}$, spread over 10 potential wells.  We note that this simple one-dimensional estimate is optimistic for the traditional Stark decelerator, as it neglects coupling between longitudinal and transverse motions and other loss mechanisms at slow forward speeds~\cite{Meerakker2006}, but it should be relatively accurate for the traveling-wave case.  Intriguingly, the results are comparable with both Zeeman-Sisyphus deceleration and direct laser cooling, though we stress again that the short pulse produced by our source is crucial for obtaining such high numbers accepted into the Stark decelerator(s).

Finally, we consider direct laser slowing from a ``two-stage'' buffer-gas cell~\cite{Patterson2007}.  Relative to the molecular beam from a single-stage source, two-stage sources produce slower beams at the expense of molecular flux. In~\cite{Hemmerling2016}, the authors report a beam of 10$^{9}$ molecules/steradian/shot, some two orders of magnitude lower than a single-stage beam, but with the mean velocity reduced to about 60~m/s.  Using direct laser slowing over a 50~cm length, the authors show that about 20\% of the beam can be slowed below 50~m/s, corresponding to about 1.5$\times$10$^{4}$ slow molecules passing through a 5~mm diameter detector located 50~cm from the source.  This number is less than the estimates above, but improvements to the molecular beam source, the effectiveness of the laser cooling, or a shortening of the source-to-detector distance, could produce significantly more slow molecules.

To summarize, we see that a number of techniques can slow CaF molecules to low velocities, and that they can have similar efficiencies. Other options not evaluated here, but certainly worthy of consideration, are the Zeeman~\cite{Narevicius2008} and centrifuge~\cite{Chervenkov2014} deceleration methods. The Zeeman-Sisyphus decelerator is competitive with other methods in terms of efficiency, does not require the exceptional branching ratios needed for direct laser slowing, and works with long, or even continuous molecular pulses that are not well suited to time-dependent deceleration methods such as used in Stark or Zeeman decelerators.

\section{Simultaneous Slowing and Cooling}
\label{Advanced_Methods}
As described in Sec.~\ref{Zeeman_Sisyphus_Deceleration}, the spread of longitudinal velocities increases as the molecules are decelerated. This is a natural consequence of a constant deceleration over a fixed decelerator length. The simulations reveal that the spread of velocities actually increases more rapidly than expected from this simple picture, especially for low laser powers.  This is because the optical pumping efficiency is greater for the slower molecules, which spend more time near the resonance points ($\bar{n}$ in equation~(\ref{PhotonsScattered}) scales as $1/v_{z}$), and so the mean deceleration is larger for slow molecules than for fast ones. In addition, because of the Doppler shift of the counter-propagating light, slower molecules must climb further up the potential energy hills to come into resonance.  Again, this results in slower molecules experiencing more average deceleration. In this section, we consider some alterations to the design of the decelerator that can minimize, or even reverse, the spread of velocities. In this way, we aim to cool and decelerate the molecules simultaneously.

\begin{figure}[!htb]
\centering
\includegraphics[width=\linewidth]{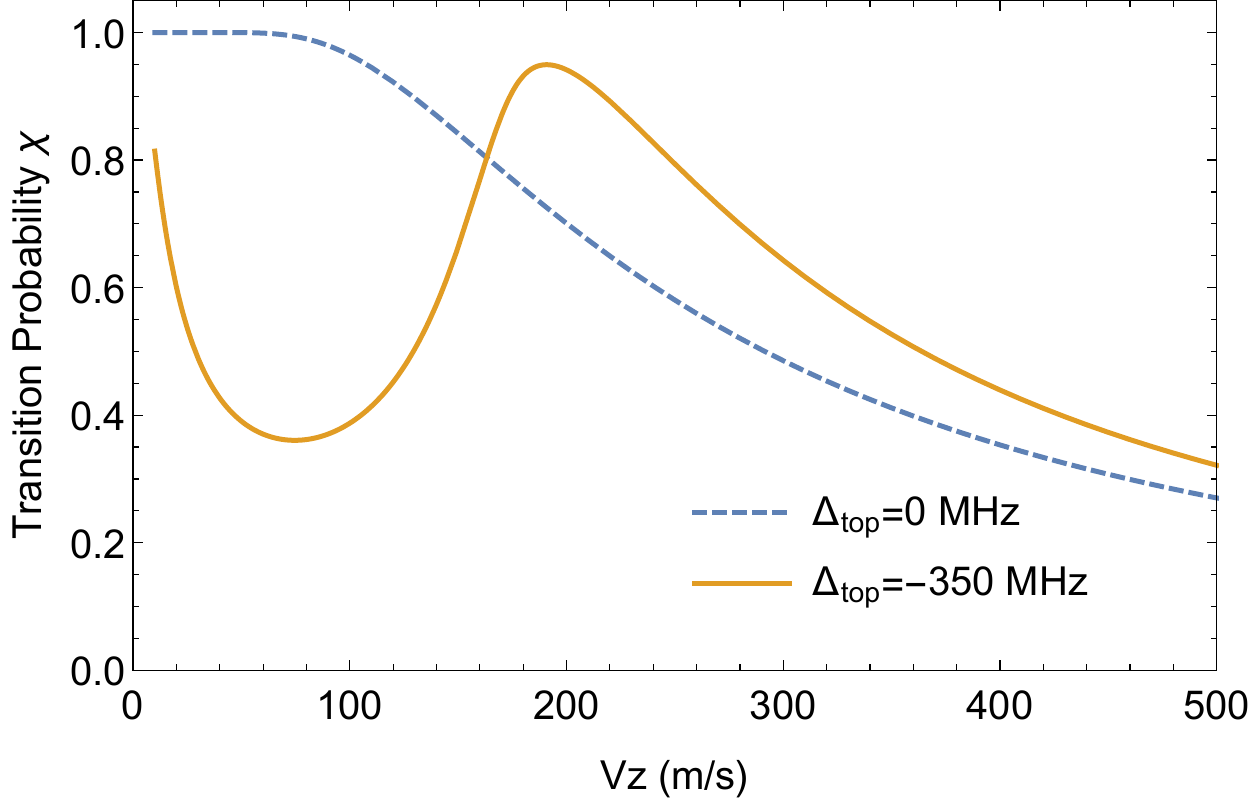}
\caption{Detuning the $L_{w \rightarrow s}$ pump laser above the potential-energy hilltop (at 14~GHz) and lowering the pump laser power (to 1~mW in this case) creates a non-monotonic velocity-selective transition probability and corresponding deceleration.}
\label{Tuning_Above_Hilltop}
\end{figure}

We first consider how to use the Doppler shift to introduce a non-monotonic velocity-dependent component to the force. To achieve this, we detune $L_{w \rightarrow s}$ above the potential-energy hilltop so that the fast molecules are Doppler-shifted into resonance at the hilltop and are optically pumped with high probability, but the slower ones are not.  We also find it necessary to reduce the $L_{w \rightarrow s}$ power to just a few mW, so that the velocity-dependent effect is not washed out by power broadening. Fortunately, those molecules that fail to optically pump are left in weak-field seeking states and so are still guided through the decelerator.  This means that the transverse stability is not adversely affected, though the decelerator does need to be made longer because of the frequent optical pumping failures.

We introduce the quantity $\Delta_{\rm top}$, the detuning of the light from resonance for a stationary molecule at the top of the potential hill. Figure~\ref{Tuning_Above_Hilltop} shows the transition probability, $\chi$, as a function of $v_{z}$ for 1~mW of power and two choices of detuning, $\Delta_{\rm top}$=0 and -350~MHz.  The value of $\chi$ is calculated by integrating the scattering rate as a molecule climbs over the top of the hill. The dashed line shows the monotonic velocity-dependence of $\chi$, and therefore also the force, in the case where $L_{w \rightarrow s}$ is tuned to the potential energy hilltop, i.e. $\Delta_{\rm top}$=0.  By contrast, when $\Delta_{\rm top}$=-350~MHz (solid line), the optical pumping efficiency and associated force is larger for faster molecules, as desired.  At very low velocities, the transition probability again increases as molecules spend a long time near the top of the potential energy hill.

We would like to compensate the changing Doppler shift as the molecules slow down, by changing $\Delta_{\rm top}$. We wish to maintain the time-independence of the deceleration process, so instead of chirping the laser frequency in time, we introduce a scaling of the magnetic field so that the magnitude at the hilltops increases with $z$. In this way, the solid curve shown in Figure~\ref{Tuning_Above_Hilltop} will be swept inwards towards lower velocities, bunching molecules in velocity as they proceed through the decelerator. The faster molecules are Doppler-shifted into resonance with the light throughout the decelerator, while the slower ones join the deceleration process later on. This is similar to the traditional Zeeman slower for atoms~\cite{Phillips1982}.

\begin{figure*}[!htb]
\centering
\includegraphics[width=\linewidth]{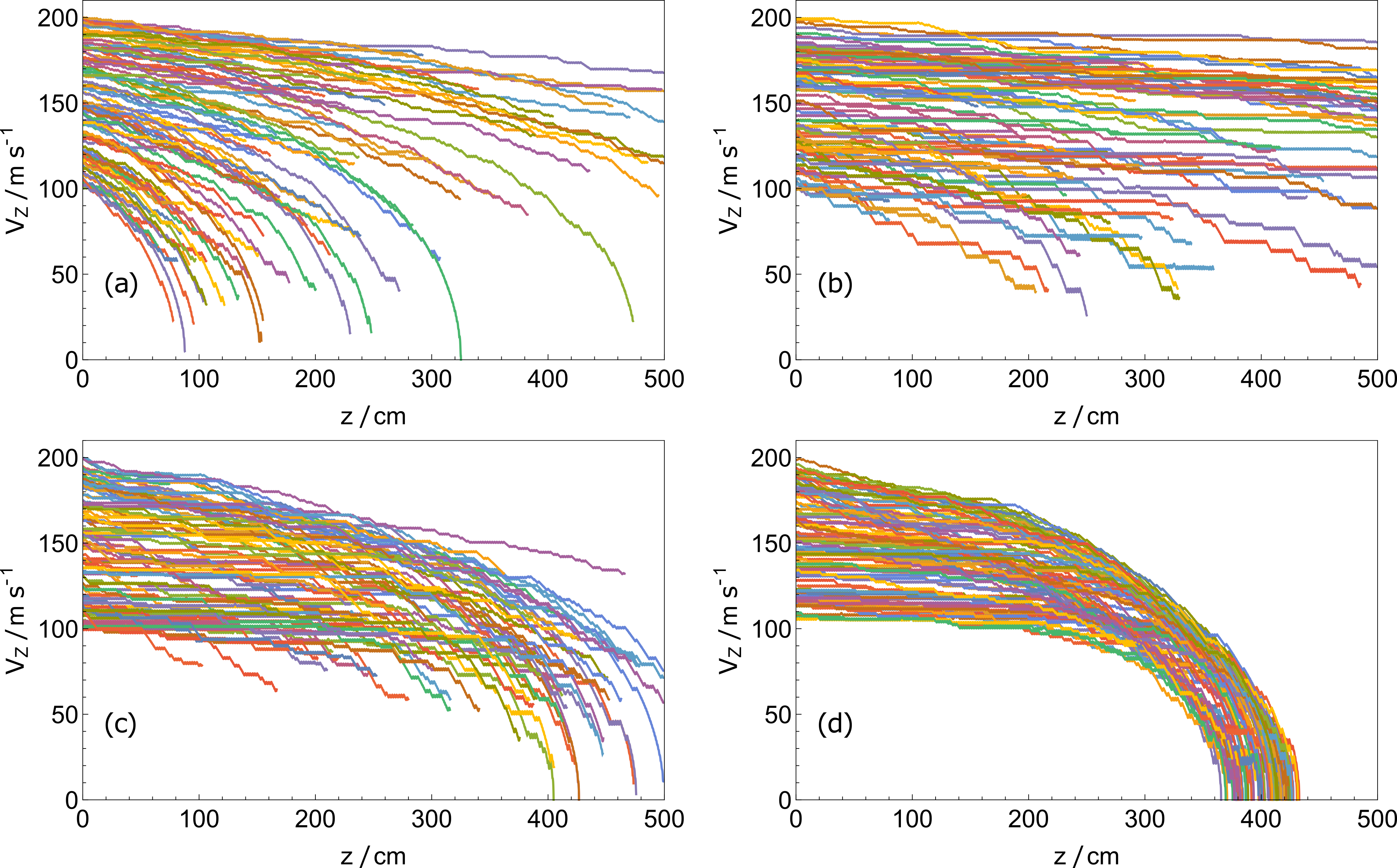}
\caption{Forward velocity versus position in the decelerator, using the refinements discussed in Sec.~\ref{Advanced_Methods}. The powers of $L_{s \rightarrow w}$ and $L_{w \rightarrow s}$ are 200~mW and 3~mW respectively, and we set $\Delta_{s \rightarrow w}$=2~GHz. (a) $\Delta_{\rm top}$=0. (b) $\Delta_{\rm top}$=-300~MHz. (c) $\Delta_{\rm top}$=-300~MHz plus a longitudinal scaling of the magnetic field amplitude of  1+0.001$z^{2}$. (d) Same as (c) but with molecules restricted to the decelerator axis.}
\label{Tuning_Above_Hilltop_Sims}
\end{figure*}

This mechanism of velocity compression is inhibited by any effect that makes $\Delta_{\rm top}$ inhomogeneous. This includes the hyperfine structure, the different Zeeman shifts of molecules at different radial positions at the hilltop, and the Zeeman shift of the excited state. The last of these is relevant because some of the ground-state sublevels couple only to the wfs manifold of the excited state, while others couple only to the sfs manifold.  We find that for CaF, the upper state Zeeman splitting is the biggest concern, being $\approx$600~MHz at the 1~T hilltops, which is 3.6 times larger than the Doppler shift of a 100~m/s molecule. A possible solution to this problem is to couple together the two excited state manifolds using an rf magnetic field tuned to the Zeeman splitting of the excited state at the hilltop. In this way, all the lower levels can couple to the lowest energy manifold of the excited state and the problem of the excited state Zeeman splitting is eliminated. The field uniformity at the hilltop is also a concern, though this could easily be improved with some minor adjustments to the wedge magnet array~\cite{Jardine2001}. To investigate the basic mechanism of the velocity compression without these complications, we set the upper state Zeeman shift to zero and limit the initial transverse distribution to be 1~mm (FWHM) and 1~m/s (FWHM).

Figure \ref{Tuning_Above_Hilltop_Sims} shows how molecules with a broad range of initial speeds propagate through the refined decelerator for four different conditions.  In Figure~\ref{Tuning_Above_Hilltop_Sims}(a), $L_{w \rightarrow s}$ is detuned to bring wfs molecules into resonance right at the potential energy hilltop ($\Delta_{\rm top}$=0).  The velocity spread increases enormously as the molecules propagate through the decelerator. This is the same effect seen in Figure~\ref{Simulations_Trajectories}(b) but amplified by the lower $L_{w \rightarrow s}$ power, which is only 3~mW, and the broader initial velocity distribution for these simulations. In Figure~\ref{Tuning_Above_Hilltop_Sims}(b), $L_{w \rightarrow s}$ is detuned above the hilltop ($\Delta_{\rm top}$=-300~MHz). In this case, the transition probability resembles the solid curve in Figure~\ref{Tuning_Above_Hilltop} and the molecules are not decelerated efficiently. Figure~\ref{Tuning_Above_Hilltop_Sims}(c) is identical except that the magnetic field amplitude is multiplied by the scaling 1+0.001$z^{2}$.  This brings the fastest molecules into the slowing cycle before the slower ones.  We see that this strategy counteracts the increase in the velocity spread, even slightly reducing it.  Figure~\ref{Tuning_Above_Hilltop_Sims}(d) shows this strategy again, but with the molecules restricted to the decelerator axis.  With all molecules experiencing the same magnetic field at the hilltop, the effects are much clearer.  Fast molecules decelerate more efficiently, while slow ones don't decelerate until the magnetic field scaling brings them into resonance with the pump light.  The result is a substantial compression of the longitudinal velocity distribution during the deceleration process.  Similar results should be attainable without the restriction to the decelerator axis, by improving the field uniformity in the strong-field regions.

\section{Conclusions}
\label{Conclusions}

We have discussed in detail the principles and design of a Zeeman-Sisyphus decelerator and presented several advantages over other methods. Because it is time-independent, it is applicable to continuous beams or long molecular pulses such as those typically emitted by cryogenic buffer gas sources. A molecule such as CaF, emitted from such a source, can be brought to rest after scattering just a few hundred photons. It follows that molecules whose vibrational branching ratios are not so favorable for direct laser cooling could still be decelerated using this technique, without needing too many repump lasers. With our magnetic field design, molecules are simultaneously guided and decelerated. This is an advantage over direct laser slowing where many molecules are lost due to the ever-increasing divergence of the slowed beam. Our simulations suggest that, for CaF, the efficiency of Zeeman-Sisyphus deceleration is comparable to direct laser slowing. For heavier molecules, or those where the photon scattering rate is lower, the slowing requires a longer distance and beam divergence can be particularly problematic. In these cases, the decelerator may provide a better way to load molecular MOTs. The decelerator uses only static magnetic fields and should be relatively straightforward to construct using readily available permanent magnets.

Our simulations with CaF use the real level structure, Zeeman shifts and transition intensities in the molecule, the full 3D magnetic field map of a realistic magnet array, and a realistic laser intensity distribution. These details introduce some subtle and important effects, but the deceleration dynamics remain similar to those expected from the very simple model presented in Figure~\ref{basicIdeaFig}.
While we have analyzed only the case of CaF in detail, it seems likely that the method will be applicable to a wide range of molecules. In Sec.~\ref{CaF}, we identified some potential pitfalls that are not problematic for CaF but might be for other species, and we recommend an analysis of the particular level structure and state couplings involved in the optical pumping transition for each case of interest. Heavier diatomics, for example, usually have smaller rotational splittings.  This can be a particular concern if the $g$-factor in the excited state is not small; other transitions out of $X$($N$=1) coming into resonance in the magnetic fields present in the decelerator may violate the cycling transition requirement and necessitate repump lasers to re-introduce leaked molecules back into the optical pumping cycle.

We have shown that, with some refinements, the Zeeman-Sisyphus decelerator could compress the velocity distribution of the molecular beam during deceleration. That would make it an especially powerful new tool for producing cooled molecular beams at low speed.

Data underlying this article can be accessed from~\cite{zsData}.


\section{Acknowledgements}
\label{Acknowledgements}
We are grateful to Ben Sauer and Ed Hinds for valuable discussions. The research leading to these results has received funding from EPSRC under grants EP/I012044 and EP/M027716, and from the European Research Council under the European Union's Seventh Framework Programme (FP7/2007-2013) / ERC grant agreement 320789.


\section{Appendix: Justification of simplifying assumptions made in the trajectory simulations}
\label{Appendix}

Our trajectory simulations assume that the transition strengths and branching ratios between the various sublevels are constant throughout the decelerator. To test this assumption, we calculated the state couplings over a wide range of magnetic field magnitudes and found that they are nearly constant for all fields above about 30~mT. While the branching ratios depend only on the field magnitude, the excitation strengths also depend on the direction of the magnetic field relative to the laser polarization. The optical pumping from wfs to sfs states occurs in the strong field region where the field direction is uniform.  However, pumping from sfs to wfs states occurs in relatively low fields ($\approx$0.2~T) where the magnetic field direction is more variable.  To explore this, we used the trajectory simulations to record the magnitude and direction of the local magnetic field each time a molecule scattered a photon.  We found that the field direction is fairly uniform even for pumping from sfs to wfs states.  Specifically, the magnetic field at the resonant points is restricted to the $xy$-plane, and is centered on the $\pm\hat{x}$ axis (the strong-field directions) with an angular extent of $\pm$45$^{\circ}$.  Calculating the transition intensities over this range of magnetic field directions, with the laser polarization fixed along $\hat{y}$,  results in variations of only a few percent, justifying our approximation of constant couplings.  The numerical values for the transition intensities and branching ratios used in the simulations appear in Tables~\ref{Transition_Intensities_Table} and \ref{Branching_Ratios_Table}, respectively.

\begin{table*}[!htb]
\centering
\footnotesize
\begin{tabular}{c|cccccccccccc}
 & (1$^{-}$,1) & (1$^{-}$,0) & (1$^{-}$,-1) & (0,0) & (1$^{+}$,-1) & (1$^{+}$,0) & (1$^{+}$,1) & (2,-2) & (2,-1) & (2,0) & (2,1) & (2,2)  \\ \hline
(1,-1)$^{*}$ & 0 & 0.0003 & 0 & 0.0833 & 0 & 0.1656 & 0 & 0.0833 & 0 & 0.0008 & 0 & 0 \\
(1,0)$^{*}$ & 0.0835 & 0 & 0.0830 & 0 & 0.0003 & 0 & 0.0007 & 0 & 0 & 0 & 0.1657 & 0 \\
(1,1)$^{*}$ & 0 & 0.1662 & 0 & 0.0007 & 0 & 0.0004 & 0 & 0 & 0 & 0.0827 & 0 & 0.0833 \\
(0,0)$^{*}$ & 0 & 0 & 0.0007 & 0 & 0.1662 & 0 & 0.0829 & 0 & 0.0831 & 0 & 0.0004 & 0\\
\end{tabular}
\caption{Transition intensities used in the simulations, calculated in a 1~T magnetic field with linearly polarized light polarized 90$^{\circ}$ with respect to the strong-field direction.  Values less than 1$\times$10$^{-4}$ are shown as zero.}
\label{Transition_Intensities_Table}
\end{table*}

\begin{table*}[!htb]
\centering
\footnotesize
\begin{tabular}{c|cccccccccccc}
 & (1$^{-}$,1) & (1$^{-}$,0) & (1$^{-}$,-1) & (0,0) & (1$^{+}$,-1) & (1$^{+}$,0) & (1$^{+}$,1) & (2,-2) & (2,-1) & (2,0) & (2,1) & (2,2)  \\ \hline
(1,-1)$^{*}$ & 0 & 0.001 & 0 & 0.167 & 0 & 0.331 & 0 & 0.167 & 0.333 & 0.002 & 0 & 0 \\
(1,0)$^{*}$ & 0.167 & 0 & 0.166 & 0 & 0.001 & 0.002 & 0.001 & 0 & 0 & 0.332 & 0.331 & 0 \\
(1,1)$^{*}$ & 0.333 & 0.332 & 0 & 0.001 & 0 & 0.001 & 0 & 0 & 0 & 0.165 & 0 & 0.167 \\
(0,0)$^{*}$ & 0 & 0.001 & 0.001 & 0.332 & 0.332 & 0 & 0.166 & 0 & 0.166 & 0 & 0.001 & 0 \\
\end{tabular}
\caption{Branching ratios used in the simulation, calculated in a 1~T magnetic field.  Values less than 1$\times$10$^{-3}$ are shown as zero.}
\label{Branching_Ratios_Table}
\end{table*}

In addition to assuming constant transition strengths and branching ratios everywhere in the decelerator, the simulations also assume linear Zeeman shifts for all molecular states of interest.  This approximation holds as long as molecules do not experience spatial regions where the magnetic field strength is below 100~mT or so, as shown in Figure~\ref{CaF_Zeeman}.  According to the finite-element model of the magnetic fields present in the decelerator, the field amplitude only drops below this value within $\sim$10~$\mu$m of the (on-axis) $K$=6 guiding-stage centers.  These regions constitute only a few parts per billion of the internal decelerator volume where the molecules propagate and thus this assumption holds for nearly all molecular trajectories.

\begin{thebibliography}{10}

\bibitem{Krems}
R.~V. Krems, W.~C. Stwalley, and B.~Friedrich.
\newblock {\em Cold molecules: theory, experiment, applications}.
\newblock CRC Press, 2009.

\bibitem{Carr2009}
L.~D. Carr, D.~DeMille, R.~V. Krems, and J.~Ye.
\newblock Cold and ultracold molecules: science, technology, and applications.
\newblock {\em New J. Phys.}, 11:055049, 2009.

\bibitem{DeMille2002}
D.~DeMille.
\newblock Quantum computation with trapped polar molecules.
\newblock {\em Phys. Rev. Lett.}, 88:067901, 2002.

\bibitem{Hudson2006}
E.~R. Hudson, H.~J. Lewandowski, B.~C. Sawyer, and J.~Ye.
\newblock Cold molecule spectroscopy for constraining the evolution of the fine
  structure constant.
\newblock {\em Phys. Rev. Lett.}, 96:143004, 2006.

\bibitem{DeMille2008}
D.~DeMille, S.~B. Cahn, D.~Murphree, D.~A. Rahmlow, and M.~G. Kozlov.
\newblock Using molecules to measure nuclear spin-dependent parity violation.
\newblock {\em Phys. Rev. Lett.}, 100:023003, 2008.

\bibitem{Isaev2010}
T.~A. Isaev, S.~Hoekstra, and R.~Berger.
\newblock Laser-cooled {RaF} as a promising candidate to measure molecular
  parity violation.
\newblock {\em Phys. Rev. A}, 82:052521, 2010.

\bibitem{Hudson2011}
J.~J. Hudson, D.~M. Kara, I.~J. Smallman, B.~E. Sauer, M.~R. Tarbutt, and E.~A.
  Hinds.
\newblock Improved measurement of the shape of the electron.
\newblock {\em Nature}, 473:493, 2011.

\bibitem{Truppe2013}
S.~Truppe, R.~J. Hendricks, S.~K. Tokunaga, H.~J. Lewandowski, M.~G. Kozlov,
  C.~Henkel, E.~A. Hinds, and M.~R. Tarbutt.
\newblock A serach for varying fundamental constants using hertz-level
  frequency measurements of cold {CH}.
\newblock {\em Nat. Commun.}, 4:2600, 2013.

\bibitem{Eckel2013}
S.~Eckel, P.~Hamilton, E.~Kirilov, H.~W. Smith, and D.~DeMille.
\newblock Search for the electron electric dipole moment using omega-doublet
  levels in {PbO}.
\newblock {\em Phys. Rev. A}, 87:052130, 2013.

\bibitem{Baron2014}
J.~Baron, W.~C. Campbell, D.~DeMille, J.~M. Doyle, G.~Gabrielse, Y.~V.
  Gurevich, P.~W. Hess, N.~R. Hutzler, E.~Kirilov, I.~Kozyryev, B.~R. O'Leary,
  C.~D. Panda, M.~F. Parsons, E.~S. Petrik, B.~Spaun, A.~C. Vutha, and A.~D.
  West.
\newblock Order of magnitude smaller limit on the electric dipole moment of the
  electron.
\newblock {\em Science}, 343(6168):269--272, 2014.

\bibitem{Lee1987}
Y.~T. Lee.
\newblock Molecular-beam studies of elementary chemical processes.
\newblock {\em Science}, 236(4803):793--798, 1987.

\bibitem{Casavecchia2000}
P.~Casavecchia.
\newblock Chemical reaction dynamics with molecular beams.
\newblock {\em Rep. Prog. Phys.}, 63(3):355, 2000.

\bibitem{Krems2008}
R.~V. Krems.
\newblock Cold controlled chemistry.
\newblock {\em Phys. Chem. Chem. Phys.}, 10:4079--4092, 2008.

\bibitem{Chandler2010}
D.~W. Chandler.
\newblock Cold and ultracold molecules: {S}potlight on orbiting resonances.
\newblock {\em J. Chem. Phys.}, 132:110901, 2010.

\bibitem{Bethlem1999}
H.~L. Bethlem, G.~Berden, and G.~Meijer.
\newblock Decelerating neutral dipolar molecules.
\newblock {\em Phys. Rev. Lett.}, 83:1558, 1999.

\bibitem{Narevicius2008}
E.~Narevicius, A.~Libson, C.~Parthey, I.~Chavez, J.~Narevicius, U.~Even, and
  M.~G. Raizen.
\newblock Stopping supersonic oxygen with a series of pulsed electromagnetic
  coils: a molecular coilgun.
\newblock {\em Phys. Rev. A}, 77:051401, 2008.

\bibitem{Fulton2004}
R.~Fulton, A.~I. Bishop, and P.~R. Barker.
\newblock Optical {S}tark decelerator for molecules.
\newblock {\em Phys. Rev. Lett.}, 93:243004, 2004.

\bibitem{Chervenkov2014}
S.~Chervenkov, X.~Wu, J.~Bayerl, A.~Rohlfes, T.~Ganter, M.~Zeppenfeld, and
  G.~Rempe.
\newblock Continuous centrifuge decelerator for polar molecules.
\newblock {\em Phys. Rev. Lett.}, 112:013001, 2014.

\bibitem{Bethlem2000}
H.~L. Bethlem, G.~Berden, F.~M.~H. Crompvoets, R.~T. Jongma, A.~J.~A. van Roij,
  and G.~Meijer.
\newblock Electrostatic trapping of ammonia molecules.
\newblock {\em Nature}, 406:491--494, 2000.

\bibitem{Tokunaga2011}
S.~K. Tokunaga, W.~Skomorowski, P.~S. Zuchowski, R.~Moszynski, J.~M. Hutson,
  E.~A. Hinds, and M.~R. Tarbutt.
\newblock Prospecs for sympathetic cooling of molecules in electrostatic, ac,
  and microwave traps.
\newblock {\em Eur. Phys. J. D}, 65:141, 2011.

\bibitem{Lim2015}
J.~Lim, M.~D. Frye, J.~M. Hutson, and M.~R. Tarbutt.
\newblock Modeling sympathetic cooling of molecules by ultracold atoms.
\newblock {\em Phys. Rev. A}, 92:053419, 2015.

\bibitem{Zeppenfeld2009}
M.~Zeppenfeld, M.~Motsch, P.~W.~H. Pinkse, and G.~Rempe.
\newblock {Optoelectrical cooling of polar molecules}.
\newblock {\em Phys. Rev. A}, 80:041401, 2009.

\bibitem{Zeppenfeld2012}
M.~Zeppenfeld, B.~G. Englert, R.~Glockner, A.~Prehn, M.~Mielenz, C.~Sommer,
  D.~L. van Buuren, M.~Motsch, and G.~Rempe.
\newblock Sisyphus cooling of electrically trapped polyatomic molecules.
\newblock {\em Nature}, 491:570--573, 2012.

\bibitem{Prehn2016}
A.~Prehn, M.~Ibr\"ugger, R.~Gl\"ockner, G.~Rempe, and M.~Zeppenfeld.
\newblock Optoelectrical cooling of polar molecules to submillikelvin
  temperatures.
\newblock {\em Phys. Rev. Lett.}, 116:063005, 2016.

\bibitem{Comparat2014}
D.~Comparat.
\newblock Molecular cooling via {S}isyphus processes.
\newblock {\em Phys. Rev. A}, 89:043410, 2014.

\bibitem{Perez2013}
M.~Quintero-Perez, P.~Jansen, T.~E. Wall, J.~E. van~den Berg, S.~Hoekstra, and
  H.~L. Bethlem.
\newblock Static trapping of polar molecules in a traveling wave decelerator.
\newblock {\em Phys. Rev. Lett.}, 110:133003, 2013.

\bibitem{Stuhl2012}
B.~K. Stuhl, M.~T. Hummon, M.~Yeo, G.~Quemener, and J.~L. Bohn.
\newblock Evaporative cooling of the dipolar hydroxyl radical.
\newblock {\em Nature}, 492:396--400, 2012.

\bibitem{Shuman2010}
E.~S. Shuman, J.~F. Barry, and D.~DeMille.
\newblock Laser cooling of a diatomic molecule.
\newblock {\em Nature}, 467:820--823, 2010.

\bibitem{Hummon2013}
Matthew~T. Hummon, Mark Yeo, Benjamin~K. Stuhl, Alejandra~L. Collopy, Yong Xia,
  and Jun Ye.
\newblock Magneto-optical trapping of diatomic molecules.
\newblock {\em Phys. Rev. Lett.}, 110:143001, 2013.

\bibitem{Zhelyazkova2014}
V.~Zhelyazkova, A.~Cournol, T.~E. Wall, A.~Matsushima, J.~J. Hudson, E.~A.
  Hinds, M.~R. Tarbutt, and B.~E. Sauer.
\newblock Laser cooling and slowing of caf molecules.
\newblock {\em Phys. Rev. A}, {89}({5}), {MAY 16} 2014.

\bibitem{Truppe2016}
S.~Truppe, H.~Williams, M.~Hambach, N.~Fitch, T.~E. Wall, E.~A. Hinds, B.~E.
  Sauer, and M.~R. Tarbutt, 2016.

\bibitem{Barry2014}
J.~F. Barry, D.~J. McCarron, E.~B. Norrgard, M.~H. Steinecker, and D.~DeMille.
\newblock Magneto-optical trapping of a diatomic molecule.
\newblock {\em Nature}, 512:286--289, 2014.

\bibitem{McCarron2015}
D.~J. McCarron, E.~B. Norrgard, M.~H. Steinecker, and D.~DeMille.
\newblock Improved magneto-optical trapping of a diatomic molecule.
\newblock {\em New J. Phys.}, 17:035014, 2015.

\bibitem{Norrgard2016}
E.~B. Norrgard, D.~J. McCarron, M.~H. Steinecker, M.~R. Tarbutt, and
  D.~DeMille.
\newblock Submillikelvin dipolar molecules in a radio-frequency magneto-optical
  trap.
\newblock {\em Phys. Rev. Lett.}, 116:063004, 2016.

\bibitem{Maxwell2005}
S.~E. Maxwell, N.~Brahms, R.~deCarvalho, J.~Helton, S.~V. Nguyen, D.~Patterson,
  J.~M. Doyle, D.~R. Glenn, J.~Petricka, and D.~DeMille.
\newblock High-flux beam source for cold, slow atoms or molecules.
\newblock {\em Phys. Rev. Lett.}, 95:173201, 2005.

\bibitem{Hutzler2012}
N.~R. Hutzler, H-I Lu, and J.~M. Doyle.
\newblock The buffer gas beam: an intense, cold, and slow source for atoms and
  molecules.
\newblock {\em Chem. Rev.}, 112:4803--4827, 2012.

\bibitem{Barry2012b}
J.~F. Barry, E.~S. Shuman, E.~B. Norrgard, and D.~DeMille.
\newblock Laser radiation pressure slowing of a molecular beam.
\newblock {\em Phys. Rev. Lett.}, 108:103002, 2012.

\bibitem{Tarbutt2015}
M.~R. Tarbutt and T.~C. Steimle.
\newblock {Modeling magneto-optical trapping of CaF molecules}.
\newblock {\em Phys. Rev. A}, 92:053401, 2015.

\bibitem{Kozyryev2016}
I.~Kozyryev, L.~Baum, K.~Matsuda, B.~Hemmerling, and J.~M. Doyle, 2016.

\bibitem{Fabrikant2014}
M.~I. Fabrikant, N.~J. Fitch, N.~Farrow, T.~Li, J.~Weinstein, and H.~J.
  Lewandowski.
\newblock Traveling wave deceleration of a buffer-gas beam of {CH} molecules.
\newblock {\em Phys. Rev. A}, 90:033418, 2013.

\bibitem{DeMille2013}
D.~DeMille, J.~F. Barry, E.~R. Edwards, E.~B. Norrgard, and M.~H. Steinecker.
\newblock On the transverse confinement of radiatively slowed molecular beams.
\newblock {\em Mol. Phys.}, 111:1805--1813, 2013.

\bibitem{Breeden1981}
T.~Breeden and H.~Metcalf.
\newblock Stark acceleration of rydberg atoms in inhomogeneous electric fields.
\newblock {\em Phys. Rev. Lett.}, 47:1726, 1981.

\bibitem{Hudson2009}
E.~R. Hudson.
\newblock Deceleration of continuous molecular beams.
\newblock {\em Phys. Rev. A}, 79:061407(R), 2009.

\bibitem{Riedel2011}
J.~Riedel, S.~Hoekstra, W.~Jager, J.~J. Gilijamse, S.~Y.~T. van~de Meerakker,
  and G.~Meijer.
\newblock Accumulation of {S}tark-decelerated {NH} molecules in a magnetic
  trap.
\newblock {\em Eur. Phys. J. D}, 65:161--166, 2011.

\bibitem{Lu2014}
H-I Lu, I.~Kozyrev, B.~Hemmerling, J.~Piskorski, and J.~M. Doyle.
\newblock Magnetic trapping of molecules via optical loading and magnetic
  slowing.
\newblock {\em Phys. Rev. Lett.}, 112:113006, 2014.

\bibitem{Metcalf}
H.~J. Metcalf and P.~van~der Straten.
\newblock {\em Laser cooling and trapping}.
\newblock Springer, 2001.

\bibitem{Halbach1980}
K.~Halbach.
\newblock Design of permanent multipole magnets with oriented rare earth
  colbalt material.
\newblock {\em Nuc. Instrum. Methods}, 169:1--10, 1980.

\bibitem{Devlin2015}
J.~Devlin, M.~R. Tarbutt, D.~L. Kokkin, and T.~C. Steimle.
\newblock Measurements of the {Z}eeman effect in the $a^{2}\pi$ and
  $b^{2}\sigma^{+}$ states of calcium fluoride.
\newblock {\em J. Mol. Spec.}, 317:1--9, 2015.

\bibitem{Meerakker2006}
S.~Y.~T. van~de Meerakker, N.~Vanhaecke, H.~L. Bethlem, and G.~Meijer.
\newblock Transverse stability in a {S}tark decelerator.
\newblock {\em Phys. Rev. A}, 73:023401, 2006.

\bibitem{Osterwalder2010}
A.~Osterwalder, S.~A. Meek, G.~Hammer, H.~Haak, and G.~Meijer.
\newblock {\em Phys. Rev. A}, 81:051401, 2010.

\bibitem{Patterson2007}
D.~Patterson and J.~M. Doyle.
\newblock Bright, guided molecular beam with hydrodynamic enhancement.
\newblock {\em J. Chem. Phys.}, 126:154307, 2007.

\bibitem{Hemmerling2016}
B.~Hemmerling, E.~Chae, A.~Ravi, L.~Anderegg, G.~K. Drayna, N.~R. Hutzler,
  A.~L. Collopy, J.~Ye, W.~Ketterle, and J.~M. Doyle, 2016.

\bibitem{Phillips1982}
W.~D. Phillips and H.~Metcalf.
\newblock Laser deceleration of an atomic beam.
\newblock {\em Phys. Rev. Lett.}, 48:596, 1982.

\bibitem{Jardine2001}
A.~P. Jardine, P.~Fouquet, J.~Ellis, and W.~Allison.
\newblock {Hexapole magnet system for thermal energy $^{3}$He atom
  manipulation}.
\newblock {\em Rev. Sci. Instr.}, 72(10):3834--3841, 2001.

\bibitem{zsData}
{Supporting data are available through Zenodo and may be used under the
  Creative Commons CCZero licence.}

\end{thebibliography}

\end{document}